\documentclass[aps,10pt,showpacs,superscriptaddress]{revtex4-2}
\usepackage{hyperref}
\hypersetup{
  colorlinks=true,        
  linkcolor=blue,         
  citecolor=blue,         
  urlcolor=blue           
}

\usepackage{graphicx}
\usepackage{dcolumn}
\usepackage{bm}
\usepackage{color}
\usepackage{enumitem}
\usepackage{amsmath}
\usepackage{amssymb}
\usepackage[caption=false]{subfig}
\usepackage[capitalise]{cleveref}
\usepackage{orcidlink}
\usepackage{graphicx} 
\usepackage{array} 
\usepackage{booktabs}
\usepackage[dvipsnames]{xcolor}
\usepackage{hyperref}

\crefname{section}{Sec.}{Secs.}      
\Crefname{section}{Sec.}{Secs.}      
\crefname{Figure}{Fig.}{Figs.}      
\Crefname{Figure}{Fig.}{Figs.}      

\begin{document}

\title{Black hole solutions surrounded by an anisotropic fluid in a Kalb–Ramond two-form background}


\author{Y. Sekhmani\orcidlink{0000-0001-7448-4579}}
\email[ ]{sekhmaniyassine@gmail.com}
\affiliation{Center for Theoretical Physics, Khazar University, 41 Mehseti Street, Baku, AZ1096, Azerbaijan.}
\affiliation{Centre for Research Impact \& Outcome, Chitkara University Institute of Engineering and Technology, Chitkara University, Rajpura, 140401, Punjab, India}

\author{A. Al-Badawi\orcidlink{0000-0002-3127-3453}}
\email[]{ahmadbadawi@ahu.edu.jo
(Corresponding author)}
\affiliation{Department of Physics, Al-Hussein Bin Talal University 71111, Ma’an, Jordan}

\author{Mohsen Fathi\orcidlink{0000-0002-1602-0722}}
\email{mohsen.fathi@ucentral.cl}
\affiliation{Centro de Investigaci\'{o}n en Ciencias del Espacio y F\'{i}sica Te\'{o}rica (CICEF), Universidad Central de Chile, La Serena 1710164, Chile}

\author{A. Vachher\orcidlink{0009-0003-4991-0662}}
\email{amnishvachher22@gmail.com}
\affiliation{Centre for Theoretical Physics, 
	Jamia Millia Islamia, New Delhi 110025, India}

\author{Sushant G. Ghosh\orcidlink{0000-0002-0835-3690}}
\email{sghosh2@jmi.ac.in}
\affiliation{Centre for Theoretical Physics, Jamia Millia Islamia, New Delhi - 110 025, India}
\affiliation{Astrophysics and Cosmology Research Unit, University of KwaZulu-Natal, Durban, South Africa}

\date{\today}

\begin{abstract}
We investigate static, spherically symmetric black hole spacetimes induced by the spontaneous Lorentz-symmetry breaking of a Kalb–Ramond (KR) two-form field, non-minimally coupled to gravity, coexisting with an anisotropic fluid. By adopting a general equation of state where the radial pressure relates to the energy density via $w_1 = -1$ and the tangential pressure via an arbitrary parameter $w_2$, we derive exact analytical solutions representing black holes surrounded by diverse matter fields, including dust ($w_2=0$), radiation ($w_2=1/3$), and dark energy-like distributions ($w_2=-1/2$). A rigorous analysis of curvature invariants confirms a genuine core singularity, while the global geometry and adherence to standard energy conditions are shown to be highly sensitive to the interplay between the KR coupling ($\ell$), the fluid density parameter ($K$), and $w_2$. Furthermore, we analyze null geodesics in detail to determine the photon sphere and shadow radii. Using the Gibbons-Werner geometrical approach and the Gauss-Bonnet theorem applied to the optical metric, we compute the weak deflection angle of light and demonstrate that both the KR field and the anisotropic fluid significantly enhance light bending, particularly in dark-energy-like backgrounds. In the strong deflection limit, we calculate the lensing observables—$\theta_\infty$, $s$, and $r_{\mathrm{mag}}$—for the supermassive black holes Sgr A* and M87*. Using EHT observations, we obtain constraints on the model parameters: for dust ($w_2=0$), the data of Sgr A* restricts $0\le \ell \le 0.065$ and $0\le K \le 0.04$, while for radiation ($w_2=1/3$), $K$ lies in $0.65\le K \le 0.85$ with $\ell$ unconstrained. We also derive similar bounds from M87*. These results offer novel astrophysical signatures for constraining string-inspired KR gravity and anisotropic dark matter halos using current EHT and future ngEHT observations.
\end{abstract}

\maketitle

\section{Introduction}

Lorentz invariance stands as the foundational cornerstone of modern theoretical physics, underpinning both the Standard Model of particle physics and General Relativity. While this symmetry has withstood rigorous experimental scrutiny across diverse energy scales, various high-energy frameworks—including string theory and non-commutative field theories—postulate that it may falter at the Planck scale \cite{Kostelecky:1988zi,Jacobson:2000xp,Carroll:2001ws,Alfaro:2001rb,Dubovsky:2004ud,Cohen:2006ky,Bengochea:2008gz,Horava:2009uw}. To systematically investigate these violations, the Standard Model Extension (SME) was established as a comprehensive effective field theory \cite{Kostelecky:2003fs}. A prominent realisation of LSB within this framework is the Bumblebee model, in which a vector field (the "bumblebee" field) is non-minimally coupled to the gravitational sector. This field acquires a non-vanishing vacuum expectation value (VEV), triggering the symmetry breaking \cite{Bluhm:2008yt,Bailey:2006fd,Kostelecky:1989jp,Kostelecky:1989jw}. Notably, Casana et al. \cite{Casana:2017jkc} derived a Schwarzschild-like black hole solution within this context, sparking extensive research into the astrophysical implications of bumblebee gravity \cite{KR1,KR2,KR3,KR4,KR5,KR6,KR7,KR8,KR9,KR10,KR11,KR12,KR13,KR14,KR15,KR16,KR17,KR18,Islam:2024sph}. Alternatively, LSB can be induced via a Kalb-Ramond (KR) field, i.e., a rank-two antisymmetric tensor field. Much like the bumblebee field, the KR field can couple non-minimally to gravity and develop a non-zero VEV \cite{Altschul:2009ae}. The dynamics of KR-mediated symmetry breaking are detailed in \cite{Nair:2021hnm,Chakraborty:2016lxo,Kalb:1974yc,Kao:1996ea}, with a primary exact solution provided by \cite{Lessa:2019bgi}. Subsequent investigations have expanded on these foundations \cite{Kumar:2020hgm,Maluf:2021eyu,Lessa:2020imi}, including the derivation of alternative solutions that further probe the limits of Lorentz-violating spacetime \cite{Yang:2023wtu}.

\noindent A fundamental question in relativistic astrophysics concerns the nature of matter configurations that can maintain static equilibrium in the vicinity of a black hole. Although ordinary baryonic matter typically fails to achieve stability due to intense gravitational attraction and radiation pressure, certain field configurations, such as the electromagnetic stress-energy tensor in the Reissner-Nordström solution, demonstrate that equilibrium is possible when the pressure profile is inherently anisotropic \cite{1985PThPh..73..288T,Kim:2019ojs}. Specifically, these systems often feature a negative radial pressure $(p_r=-\rho)$, suggesting that anisotropy and non-standard equations of state are essential for understanding the coexistence of compact objects and surrounding matter \cite{Stephani:2003tm,Delgaty:1998uy,Semiz:2008ny}. Traditionally, stellar and black hole models have relied on the Pascalian (isotropic) fluid approximation, a simplification supported by broad observational data in standard Einstein gravity \cite{Katsuragawa:2015lbl,Hendi:2017ibm,Hendi:2016uni}. However, as theoretical physics moves toward modified frameworks like massive gravity or KR theory, the role of local anisotropy has become a focal point of modern research \cite{Isayev:2017hup,Pant:2014dna,Harko:2002pxr,Mak:2001eb}. Such anisotropy is not merely a mathematical curiosity; it arises naturally in self-gravitating systems with exotic thermodynamic properties or high-density regimes, such as quark stars or systems governed by barotropic equations of state \cite{Herrera:1997plx,Varela:2010mf,Bowers:1974tgi,Ivanov:2002jy}. Recent advancements, including the derivation of covariant Tolman-Oppenheimer-Volkoff (TOV) equations for non-isotropic fluids \cite{Carloni:2017bck}, have highlighted how pressure gradients significantly influence the structural stability and evolutionary trajectories of relativistic objects \cite{matese1980new,Mak:2001eb}. Building upon this momentum, the present work explores a novel black hole solution within the context of KR gravity, specifically examining the gravitational and thermodynamic implications of an environmental anisotropic fluid coupled to the KR field's non-zero vacuum expectation value.

\noindent The investigation of gravitational lensing within the strong-field limit has emerged as a robust area of inquiry, primarily because relativistic images provide a unique window into the high-curvature environment immediately surrounding an event horizon. These images encode the subtle topological and geometric nuances of the spacetime, offering a powerful diagnostic for probing gravity where it is most extreme. While modified gravity theories often converge with General Relativity (GR) in the weak-field limit, the strong-field regime serves as a critical testing ground for identifying potential departures from Einsteinian physics. Consequently, gravitational lensing in this context is an indispensable tool for distinguishing between competing gravitational frameworks. The theoretical trajectory of this field began with Darwin’s foundational study of light deflection around a Schwarzschild black hole \cite{Darwin1959TheGF}. This work laid the groundwork for Virbhadra and Ellis, who derived the definitive gravitational lens equation \cite{Virbhadra:1999nm}. Subsequently, Bozza and collaborators \cite{Bozza:2001xd} extended these analytical methods, enabling the systematic study of a diverse range of spacetimes beyond the standard Schwarzschild solution \cite{Bozza:2002zj,Feleppa:2024vdk,Bhadra:2003zs,Shaikh:2019itn}. In the present study, we utilize these analytical formalisms to detect signatures of anisotropic fluid and Lorentz Symmetry Breaking (LSB). Building on existing literature regarding anisotropic signatures \cite{perna2009gravitational} and LSB effects in specific spacetimes \cite{Junior:2024vdk,Yang:2023wtu}, we aim to quantify how these phenomena distort the path of electromagnetic radiation. To ensure our theoretical model remains physically grounded, we contrast our results with high-precision observational data. Specifically, we employ the deviation parameters $\delta$ established by observations of the Supermassive Black Holes (SMBHs) M87* \cite{EventHorizonTelescope:2022xqj,EventHorizonTelescope:2021dqv} and Sgr A* \cite{Do:2019txf,GRAVITY:2021xju}. By mapping our model's predictions against these empirical bounds, we can establish rigorous constraints on the theory's free parameters and assess its overall feasibility.

\noindent The structure of this paper is organized as follows: In Section \ref{sec2}, we derive the static, spherically symmetric black hole solutions within the proposed framework. Section \ref{sec3} is dedicated to the analysis of null geodesic trajectories, utilizing Event Horizon Telescope (EHT) data to establish rigorous constraints on the black hole parameters. Subsequently, Section \ref{sec:weak_deflection} investigates the weak deflection angle, while Sections  \ref{sec:strongdeflection} and \ref{sec:anaylsis} provide a comprehensive study of strong gravitational lensing, further refining our parameter space in light of the M87* and Sgr A* observational bounds. Finally, Section \ref{sec:conlusion} summarizes our primary findings and outlines prospective avenues for future research regarding this black hole model.

\section{BH surrounded by PFDM with a background KR field} \label{sec2}

The Lorentz Symmetry Breaking (LSB) considered in this article is induced by a non-zero vacuum expectation value (VEV) of the KR two-form $B_{\mu\nu}$, an antisymmetric tensor of rank two. The KR sector is non-minimally coupled to gravity. Our objective is to obtain static, spherically symmetric black-hole solutions in which a background KR field coexists with an \emph{anisotropic} matter distribution (henceforth the KR--anisotropic--fluid system). The total action we employ reads \cite{Kalb:1974yc,Kao:1996ea,Kar:2002xa,Altschul:2009ae,Fu:2012sa,Chakraborty:2014fva,Chakraborty:2016lxo,Lessa:2019bgi,Kumar:2020hgm,Nair:2021hnm}:
\begin{widetext}
\begin{align}\label{action_modified}
S=\int d^4x\sqrt{-g}\bigg[\frac{1}{2\kappa}\Big(R-2\Lambda+\varepsilon\, B^{\mu\lambda}B^{\nu}{}_{\lambda}R_{\mu\nu}\Big)
-\frac{1}{12}H_{\lambda\mu\nu}H^{\lambda\mu\nu}-V\big(B_{\alpha\beta}B^{\alpha\beta}\pm b^2\big)+\mathcal{L}_{\rm aniso}\bigg],
\end{align}
\end{widetext}
where $\kappa=8\pi G$ ($G$ being the Newtonian gravitational constant), $\varepsilon$ is the non-minimal coupling constant, $b^2>0$ fixes the norm of the KR VEV, and $\Lambda$ is the cosmological constant. The KR field strength is defined as $H_{\mu\nu\rho}\equiv\partial_{[\mu}B_{\nu\rho]}$. In \eqref{action_modified}, $\mathcal{L}_{\rm aniso}$ denotes the Lagrangian density of the anisotropic fluid, which constitutes the primary matter sector. The self-interaction potential $V(X)$, where $X = B_{\alpha\beta}B^{\alpha\beta}\pm b^2$, triggers spontaneous Lorentz symmetry breaking, yielding a non-zero VEV $\langle B_{\mu\nu}\rangle=b_{\mu\nu}$ subject to the constraint $b_{\mu\nu}b^{\mu\nu}=\mp b^2$. In the vacuum configuration, this constraint implies that the KR field strength vanishes in the background.

Varying the action \eqref{action_modified} with respect to $g^{\mu \nu}$ leads to the field equations:
\begin{align} 
R_{\mu \nu }-\frac{1}{2}g_{\mu \nu }R+\Lambda g_{\mu \nu }=\kappa \left( T^{\text{KR}}_{\mu\nu}+T^{\text{M}}_{\mu\nu}\right), \label{fe} 
\end{align} 
where $T^{\text{M}}_{\mu\nu}$ is the energy-momentum tensor of the anisotropic fluid and $T^{\text{KR}}_{\mu\nu}$ is the effective energy--momentum tensor of the KR field:
\begin{align} \label{tkr}
\kappa T^{\text{KR}}_{\mu\nu} &= \frac{1}{2} H_{\mu \alpha \beta } H_{\nu }{}^{\alpha \beta } - \frac{1}{12} g_{\mu \nu } H^{\alpha \beta \rho } H_{\alpha \beta \rho} + 2V'(X) B_{\alpha\mu}B^{\alpha}{}_\nu - g_{\mu\nu}V(X) \nonumber \\ 
&\quad + \varepsilon \bigg[\frac{1}{2} g_{\mu \nu } B^{\alpha \gamma } B^{\beta }{}_{\gamma }R_{\alpha \beta } - B^{\alpha }{}_{\mu } B^{\beta }{}_{\nu }R_{\alpha \beta } - B^{\alpha \beta } B_{\nu \beta } R_{\mu \alpha }-B^{\alpha \beta } B_{\mu \beta } R_{\nu \alpha } \nonumber \\ 
&\quad + \frac{1}{2} \nabla _{\alpha }\nabla _{\mu }\left(B^{\alpha \beta } B_{\nu \beta }\right) + \frac{1}{2} \nabla _{\alpha }\nabla _{\nu }\left(B^{\alpha \beta } B_{\mu \beta }\right) - \frac{1}{2}\nabla ^{\alpha }\nabla _{\alpha }\left(B_{\mu }{}^{\gamma }B_{\nu \gamma } \right) \nonumber \\ 
&\quad - \frac{1}{2} g_{\mu \nu } \nabla _{\alpha }\nabla _{\beta }\left(B^{\alpha \gamma } B^{\beta }{}_{\gamma }\right)\bigg].
\end{align} 
Here, the prime denotes the derivative with respect to the argument $X$. The Bianchi identities ensure the conservation of the combined tensor $T^{\text{KR}}_{\mu\nu}+T^{\text{M}}_{\mu\nu}$. In the subsequent analysis, we consider a vanishing cosmological constant ($\Lambda = 0$).

The anisotropic fluid contribution manifests through its energy-momentum tensor \cite{Cho:2017nhx}:
\begin{equation}
T_{\mu \nu}^{\text{M}} = (\rho+p_2)u_{\mu}u_{\nu} + (p_1-p_2)x_{\mu}x_{\nu} + p_2g_{\mu\nu}, \label{pp3}
\end{equation}
where $\rho$ is the energy density measured by a comoving observer, $u^\mu$ is the timelike four-velocity, and $x^\mu$ is a spacelike unit vector orthogonal to $u^\mu$ and the angular directions. The resulting stress--energy tensor,
\begin{equation}
T_{\mu}^{\nu} = \mathrm{diag}\big(-\rho(r),\,p_{1}(r),\,p_{2}(r),\,p_{2}(r)\big),
\end{equation}
together with a static, spherically symmetric geometry, provides a flexible framework for describing both the interiors of ultra-dense objects and nontrivial matter distributions exterior to a black hole. Allowing the radial ($p_1$) and tangential ($p_2$) pressures to differ is physically well-motivated; such anisotropy naturally arises from electromagnetic fields, scalar condensates, or tangential stresses in dark-matter halos, crucially modifying equilibrium and stability properties compared to the perfect-fluid case.

To close the system, we introduce equations of state that relate pressure to energy density. Rather than imposing a single, global barotropic index, we adopt the more general ansatz
\begin{equation}\label{gen_eos}
p_{1}(r)=w_{1}(r)\,\rho(r),\qquad p_{2}(r)=w_{2}(r)\,\rho(r),
\end{equation}
where \(w_{1}\) and \(w_{2}\) may be radial functions and need not coincide. This parametrisation simultaneously captures the standard cosmological limits and genuine anisotropic behaviours:
\begin{itemize}
  \item \textbf{Dust:} \(w_{1}=w_{2}=0\).
  \item \textbf{Radiation:} \(w_{1}=w_{2}=1/3\).
  \item \textbf{Dark energy (cosmological--constant limit):} \(w_{1}=w_{2}\simeq -1/2\).
  \item \textbf{Phantom:} \(w_{i}<-1\).
\end{itemize}
In particular, cold dark matter or PFDM profiles are naturally represented by \(w_{1}\approx0\) while \(w_{2}\) may deviate slightly from zero to encode small tangential stresses; conversely, a dominant negative radial pressure (\(w_{1}<0\)) models locally repulsive dark–energy–type behaviour.

To derive static, spherically symmetric solutions, we employ the metric ansatz:
\begin{equation}
ds^2=-F(r)dt^2+G(r)dr^2+r^2 d\theta^2+r^2 \sin^2\theta d\phi^2.
\label{trial}
\end{equation}

We consider a pseudoelectric KR gravity field configuration in which only the $b_{01}$ and $b_{10}$ components are non-zero. The constant norm condition yields:
\begin{equation}
b_{01}=-b_{10}=|b| \sqrt{\frac{G(r)F(r)}{2}}.
\end{equation}

Assuming the KR gravity field remains frozen at its VEV, the field equations reduce to:

\begin{align}
-\frac{r G'(r) + \big(G(r)-1\big)G(r)}{r^2 G(r)^2}
&= \frac{\ell}{2\,r^2 F(r)^2 G(r)^2}\,\mathcal{E}_0 - 8\pi\,\rho(r),
\label{eq:00_compact}\\[6pt]
\frac{r F'(r) + F(r)\big(1-G(r)\big)}{r^2 F(r) G(r)}
&= \frac{\ell}{2\,r^2 F(r)^2 G(r)^2}\,\mathcal{E}_1 + 8\pi\,p_1(r),
\label{eq:11_compact}\\[6pt]
\frac{\mathcal{M}}{4 r F(r)^2 G(r)^2}
&= -\,\frac{\ell}{4 r F(r)^2 G(r)^2}\,\mathcal{E}_2 + 8\pi\,p_2(r).
\label{eq:22_compact}
\end{align}
with 
\begin{align}
\mathcal{E}_0 &:= r^2 G(r)\,F'(r)^2 + r F(r)\Big(r F'(r) G'(r) - 2 G(r)\big(r F''(r)+F'(r)\big)\Big) + 2 F(r)^2 G(r),\\
\mathcal{E}_1 &:= r^2 G(r)\,F'(r)^2 + r^2 F(r)\Big(F'(r) G'(r) - 2 G(r) F''(r)\Big) + 2 F(r)^2\big(r G'(r) + G(r)\big),\\
\mathcal{E}_2 &:= r G(r)\,F'(r)^2 + F(r)\Big(r F'(r) G'(r) - 2 G(r)\big(r F''(r)+F'(r)\big)\Big) + 2 F(r)^2 G'(r),\\
\mathcal{M}   &:= -r G(r)\,F'(r)^2 + F(r)\Big(2 G(r)\big(r F''(r)+F'(r)\big) - r F'(r) G'(r)\Big) - 2 F(r)^2 G'(r).
\end{align}
Assuming \(w_1=-1\), subtracting Eq.~\eqref{eq:00_compact} from Eq.~\eqref{eq:11_compact} yields
\begin{equation}
\frac{d}{dr}\ln\big[F(r)G(r)\big]=0,
\end{equation}
hence \(F(r)G(r)=C\). Imposing the usual asymptotic normalization fixes \(C=1\), hence \(G(r)=F(r)^{-1}\). Consequently, the line element reduces to
\begin{equation}\label{eq:metric}
ds^2 = - F(r) dt^2 + F(r)^{-1} dr^2 + r^2 (d\theta^2+ \sin^2\theta \, d\phi^2).
\end{equation}
This means that there exists a hypersurface-orthogonal Killing vector in the spacetime.
Thus, the spacetime is static in the region where $f>0$, and $\rho=\rho(r)$ and $p_2 = p_2(r)$ hold by consistency.
Equation~\eqref{eq:00_compact} can be formally integrated to give
\begin{equation}
F(r) = \frac{1}{1-\ell}-\frac{2m(r)}{r},\label{ff}
\end{equation}
where the mass function $m(r)$ is defined by
\begin{equation}
 m(r) = 4\pi \int^r r'^2 \rho(r') dr'.
\end{equation}
Here, the integration constant is absorbed into the definition of $m(r)$.
If one requires the analyticity of the spacetime at the center, it requires
$m(r) \simeq  m_3 r^3 + m_5 r^5 + \cdots$ around $r=0$, where $m_3$, $m_5$ are the constants, which restricts the form of $\rho(r)$.
Putting Eq.~\eqref{ff} to Eq.~\eqref{eq:22_compact}, we obtain the expression of $p_2$ in terms of $\rho$ as
\begin{equation}
p_2 = \frac{1}{2} (\ell-1) \left(r \rho '(r)+2 \rho (r)\right),\label{EE}
\end{equation}
which can also be obtained from the conservation law $\nabla^\mu T_{\mu \nu} =0$.

The purpose of this work is to find analytic solutions of Einstein's equations.
In this work, we restrict our interests to the exactly solvable case with 
\begin{equation}
	w_1=-1.
\end{equation}
When $\rho$ plays the role of an energy density, the energy conditions restrict the matter kinds to physically allowed ones.
Among the conditions, the positivity of energy density appears to be crucial.
In addition to it, all the energy conditions require $ w_2 \geq -1$.
Specifically, the dominant energy condition requires $w_2 \leq 1$ and the strong energy condition requires $w_2 \geq 0 $.
Therefore, when $0\leq w_2 \leq 1$, all the energy conditions are satisfied.
Once we assume $p_2 = w_2 \rho$, Eq.~\eqref{EE} is solved to give $m(r)$ for $w_2 \neq 1/2$, the density and the radial pressure, 
\begin{equation}
	m(r) = M + \frac{K\, r^{1+\frac{2 w_2}{\ell-1}}}{2 (1-\ell)}, \qquad
\rho (r) = -p_1(r) = \frac{K\, (1-2 w_2-\ell) r^{\frac{2 (1+w_2-\ell)}{\ell-1}}}{8 \pi  (1-\ell)^2},\label{mp}	
\end{equation}
where $M$ and $K$ are constants.
For the energy density to be non-negative, we require
\begin{equation}
r_0^{2w_2} \equiv (1-2w_2-\ell) K \geq 0,\label{r0}
\end{equation}
where the positive parameter $r_0$ of  length (mass) scale was introduced for convenience
because the dimension of the parameter $K$ is dependent on the value of $w_2$.
The energy density and the pressure are singular at the origin or at infinity when $w_2> -1$ and $w_2< -1$, respectively.

To have a smooth $w_2 \to (1-\ell)/2$ limit, we introduce a new mass parameter 
\begin{equation}
M' \equiv M + \frac{r_0^{1-\ell}}{2(1-2w_2-\ell)}.\label{mprime}
\end{equation}
Then, the solutions for $w_2 = (1-\ell)/2$ can be specified by taking the limit $w_2\to (1-\ell)/2$ from Eq.~\eqref{mp}, which gives  
$$
m(r) = M' + \frac{r_0}{2} \log \frac{r}{r+r_0} , \qquad \rho(r) = \frac{r_0^{1-\ell}}{8\pi(1-\ell)^2\, r^3}. 
$$
All the other physical formulae for $w_2=1/2$ in this work can be obtained in the same manner.
Therefore, we will not discuss the $w_2=1/2$ case separately.  
The metric function in Eq.~\eqref{ff} becomes
\begin{equation}
 F(r) = \frac{1}{1-\ell}-\frac{2 M}{r}-\frac{K}{1-\ell}\,r^{\frac{2 w_2}{\ell-1}},\label{m.function}
\end{equation}
where $M$ and $K$ can be rewritten by using Eqs.~\eqref{r0} and \eqref{mprime}.
Because we are interested in solutions involving matter, we restrict our attention to the case with $r_0 \neq 0$.
For $1/2< w_2 \leq 1$, the spacetime structure must be very similar to that of the Reissner-Nordstr\"om geometry in coupling to a self-interacting KR field.
For \(1/2<w_2\leq1\) the anisotropic-fluid term is sufficiently short-ranged that it produces only a localized correction to the metric; as a consequence the causal and horizon structure closely resembles that of a Reissner–Nordström black hole when the geometry is coupled to a self-interacting KR two-form. In the isotropic limit \(w_1=w_2=-1\), the combinations \(M/(1-\ell)\) and \(3K/(1-\ell)\) play the roles of an effective mass parameter and an effective cosmological term within the KR framework — and, for those values of \(\ell\) that make the exponent \(2w_2/(\ell-1)\) produce an \(r^2\) scaling, the \(K/(1-\ell)\) contribution reduces to an (anti-)de Sitter cosmological constant while \(M/(1-\ell)\) is the ADM mass.

\begin{figure}[ht!]
    \centering
    \begin{tabular}{ccc}
        \includegraphics[width=0.32\textwidth]{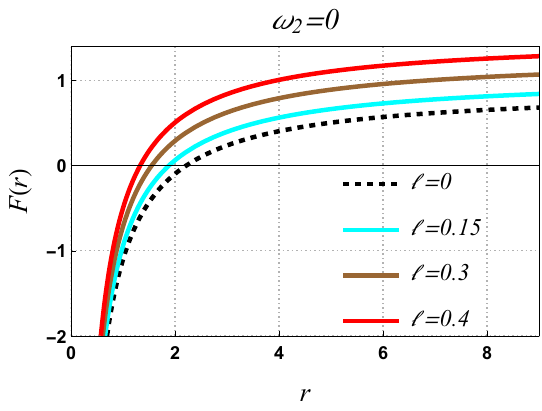} & 
        \includegraphics[width=0.32\textwidth]{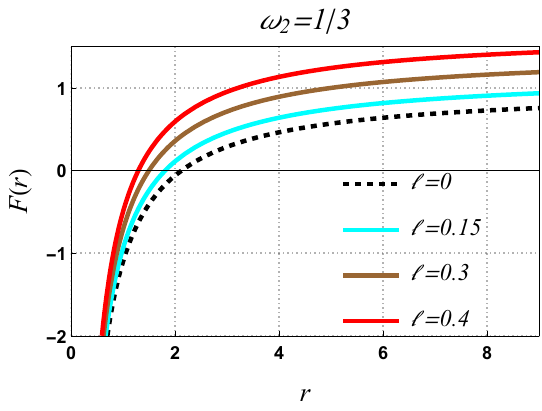} & 
        \includegraphics[width=0.32\textwidth]{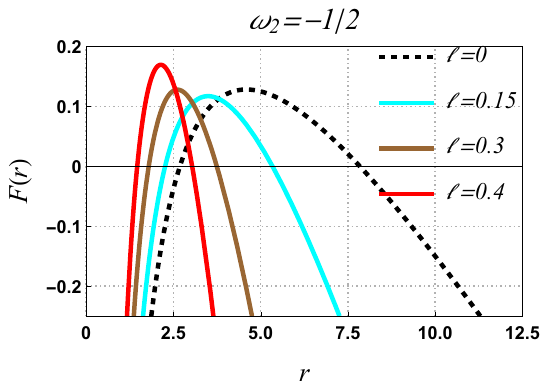} \\
    \end{tabular}
    \caption{\footnotesize $F(r)$ metric function with $K=0.095$ and $M=1$.}
    \label{frpr}
\end{figure}

In Fig.~\ref{frpr} we display the behaviour of the metric function $f(r)$ for \(M=1\) and representative values of the KR coupling \(\ell\) and the anisotropy parameter \(K\). Two correlated effects are visible. First, the overall amplitude of \(f(r)\) is raised as \(\ell\) approaches unity because of the global prefactor \(1/(1-\ell)\), which shifts the baseline and therefore moves the radial locations of simple zeros (horizons) systematically outward. Second, the anisotropic-fluid term \(-K\,r^{2w_{2}/(\ell-1)}\) changes its radial scaling depending on the equation-of-state parameter \(w_{2}\): for \(w_{2}=0\) (dust) it reduces to a constant offset that primarily translates the curve without altering asymptotic decay; for \(w_{2}=1/3\) (radiation) it decays at large \(r\), producing a Reissner–Nordström–like inner/outer-root profile; and for \(w_{2}<0\) (dark-energy–like) it grows with \(r\), generating a cosmological-type rise which can introduce an additional cosmological root. Hence, Fig.~\ref{frpr} compactly illustrates how the interplay between the \(\ell\)-controlled amplitude and the sign of the exponent \(2w_{2}/(\ell-1)\) determines both the number of horizons and their radial positions across the parameter sweep.

\textbf{}

Now, let us analyze the curvature singularities. The scalar curvature reads
\begin{equation}
R = \frac{2 (\ell+w_2-1) r_0^{2 w_2} r^{\frac{2 w_2}{\ell-1}} + 2 (1-\ell)^2 \ell}{(\ell-1)^3 r^2},
\end{equation}
is generically singular at the origin due to the $r^{-2}$ term. At infinity, it decays only when $w_2/(\ell-1) < 1$; if $w_2/(\ell-1) > 1$, the curvature diverges as $r \to \infty$, indicating a non-asymptotically flat geometry. The squared Ricci tensor is given by
\begin{equation}
R_{ab}R^{ab} = \frac{2 (1-\ell)^4 \ell^2 + 4 (\ell-1)^3 \ell r_0^{2 w_2} r^{\frac{2 w_2}{\ell-1}} + 2 \left((1-\ell)^2 + w_2^2\right) r_0^{4 w_2} r^{\frac{4 w_2}{\ell-1}}}{(\ell-1)^6 r^4},
\end{equation}
which diverges at the origin as $r^{-4}$ (or more severely if $w_2/(\ell-1) < 0$). Similar to the Ricci scalar, it decays at infinity only for $w_2/(\ell-1) < 1$. 

The Kretschmann invariant for this metric is given by
\begin{align}
R_{abcd}R^{abcd} &= \frac{4}{r^6}\Bigg(
\frac{4\big((\ell-1)^2 M(\ell+2w_2-1)-w_2\,r_0^{2w_2}\,r^{\frac{2w_2}{\ell-1}+1}\big)^2}
{(1-\ell)^4(\ell+2w_2-1)^2}+\left(2M+\frac{r\!}{\ell-1}\left(\dfrac{r_0^{2w_2}\,r^{\frac{2w_2}{\ell-1}}}{\ell+2w_2-1}+\ell\right)\right)^{\!2} \nonumber\\[6pt]
&\qquad +\;
\frac{\big(2(\ell-1)^3 M(\ell+2w_2-1)+w_2(-\ell+2w_2+1)\,r_0^{2w_2}\,r^{\frac{2w_2}{\ell-1}+1}\big)^2}
{(1-\ell)^6(\ell+2w_2-1)^2}\Bigg).
\end{align}
As $r \to 0$, the scalar generically diverges as $r^{-6}$ (assuming $M \neq 0$), while its asymptotic behavior at large $r$ is determined by the ratio $2w_2/(\ell-1)$. Specifically, the invariant decays for $2w_2/(\ell-1) < 2$ and approaches a non-zero limit or grows for $2w_2/(\ell-1) \ge 2$. On the other hand, the factors of $(\ell+2w_2-1)$ in the denominators appear to suggest a divergence, at $w_2 = (1-\ell)/2$; however, this is a coordinate-dependent artifact that is removable via the reparameterization \eqref{mprime}, and thus does not constitute a physical curvature divergence. For $r_0 \neq 0$, the invariant remains regular everywhere only in the specific case $M=0$ and $w_2=-1$, which corresponds to (anti-)de Sitter space. In other parameter regimes, the curvature singularity, which is located at the origin or at infinity depending on the sign of $2w_2/(\ell-1)$, may be present, manifesting as either a black hole singularity hidden by horizons or a naked singularity, contingent upon the global horizon structure.

Energy conditions furnish indispensable criteria for assessing the physical admissibility of spacetime solutions and are routinely invoked in studies of both cosmological models \cite{
Nojiri:2010wj} and black hole geometries \cite{Hawking:1973uf,Wald:1984rg}. These conditions constrain the stress-energy tensor $T_{\mu\nu}$ in a manner that remains physically robust across Einstein gravity \cite{1915SPAW.......844E} and its various modifications \cite{Capozziello:2011et}. The standard set—comprising the Null (NEC), Weak (WEC), Strong (SEC), and Dominant (DEC) energy conditions—is defined in terms of the energy density $\rho$ and principal pressures $P_i$ as follows:

\begin{alignat}{2}
\text{\textbf{NEC}} &: \quad \rho + P_i \geq 0 \label{18} \\
\text{\textbf{WEC}} &: \quad \rho \geq 0, \quad &&\rho + P_i \geq 0 \label{19} \\
\text{\textbf{SEC}} &: \quad \rho + \sum_i P_i \geq 0, \quad &&\rho + P_i \geq 0 \label{20} \\
\text{\textbf{DEC}} &: \quad \rho \geq 0, \quad &&\rho \geq |P_i| \label{21}
\end{alignat}
Relevant quantities are deduced as
\begin{align}
   \rho + p_1 =0,\,\, \rho+ p_{2,3}&=\frac{K(w_2+1)  (1-2 w_2-\ell) r^{\frac{2 w_2}{\ell-1}-2}}{8 \pi  (1-\ell)^2},\\
   \rho+p_1+p_2+p_3&=\frac{w_2 K (1-2 w_2-\ell) r^{\frac{2 (w_2-\ell+1)}{\ell-1}}}{4 \pi  (1-\ell)^2}\,,\\
   \rho-|p_1|=0,\,\,  \rho-|p_{2,3}|&=\rho-\Bigl|\frac{w_2\, K (1-2 w_2-\ell) r^{\frac{2 (w_2-\ell+1)}{\ell-1}}}{8 \pi  (1-\ell)^2}\Bigr|\,.
\end{align}

\begin{itemize}
        \item The Null Energy Condition (NEC) is saturated, $\rho+p_1=0$,
while the tangential null combination is given by
\[
\rho+p_{2,3}=\frac{(1+w_2)\,K(1-2w_2-\ell)\,r^{\frac{2w_2}{\ell-1}-2}}{8\pi(1-\ell)^2}.
\]
Therefore, the NEC is algebraically equivalent to
\[
(1+w_2)\,K(1-2w_2-\ell) \ge 0.
\]
In particular:
\begin{itemize}
  \item \(w_2=-1\) saturates the tangential NEC.
  \item For \(w_2>-1\) the NEC reduces to \(K(1-2w_2-\ell)\ge0\).
  \item For \(w_2<-1\) the NEC requires \(K(1-2w_2-\ell)\le0\).
\end{itemize}
\end{itemize}

\begin{itemize}
   \item Using the displayed expression
\[
\rho+\sum_i p_i=\frac{w_2\,K(1-2w_2-\ell)\,r^{\frac{2(w_2-\ell+1)}{\ell-1}}}{4\pi(1-\ell)^2},
\]
the SEC (both \(\rho+\sum_i p_i\ge0\) and \(\rho+p_i\ge0\)) is equivalent to the pair of algebraic inequalities
\begin{equation}
w_2\,K(1-2w_2-\ell)\ge0\quad\text{and}\quad (1+w_2)\,K(1-2w_2-\ell)\ge0.
\end{equation}
Consequences by \(w_2\)-region:
\begin{itemize}
  \item If \(w_2>0\) then \(w_2>0\) and \(1+w_2>0\), so the SEC reduces to \(K(1-2w_2-\ell)\ge0\).
  \item If \(-1<w_2<0\) then \(w_2<0\) while \(1+w_2>0\); the two inequalities are incompatible unless \(K(1-2w_2-\ell)=0\). Hence, the SEC is generically violated for \(-1<w_2<0\).
  \item If \(w_2<-1\) both \(w_2\) and \(1+w_2\) are negative and the SEC can hold for \(K(1-2w_2-\ell)\le0\).
\end{itemize}
\end{itemize}

\begin{itemize}
   \item The radial DEC is saturated because \(p_1=-\rho\) gives \(\rho-|p_1|=0\). The tangential DEC condition is
\begin{equation}
\rho-|p_{2,3}|=\rho-\Biggl|\frac{w_2\,K(1-2w_2-\ell)\,r^{\frac{2(w_2-\ell+1)}{\ell-1}}}{8\pi(1-\ell)^2}\Biggr|\ge0.
\end{equation}
For the solution family where \(\rho\) and \(p_{2,3}\) share the same overall parameter factor \(S\) and the same radial scaling, the DEC reduces to the simple algebraic requirements
\begin{equation}
K(1-2w_2-\ell)\ge0\qquad\text{and}\qquad |w_2|\le 1.
\end{equation}
Thus the DEC requires \(K(1-2w_2-\ell)\ge0\) together with \(-1\le w_2\le1\). (If the radial exponents differed, the condition would become radius-dependent and could hold only inside a restricted radial domain.)
\end{itemize}

\begin{figure}[ht!]
    \centering
    \begin{tabular}{cc}
        \includegraphics[width=0.48\textwidth]{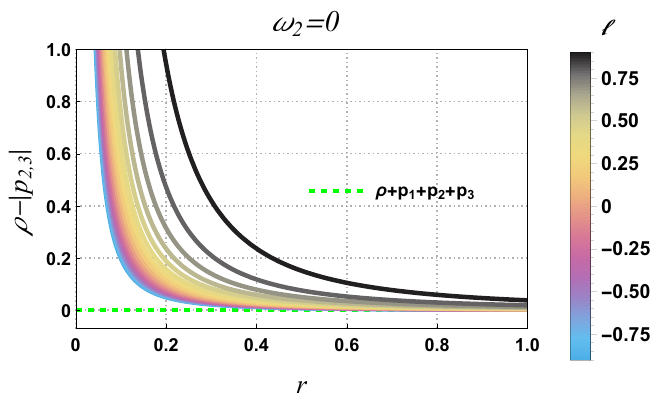} & 
        \includegraphics[width=0.48\textwidth]{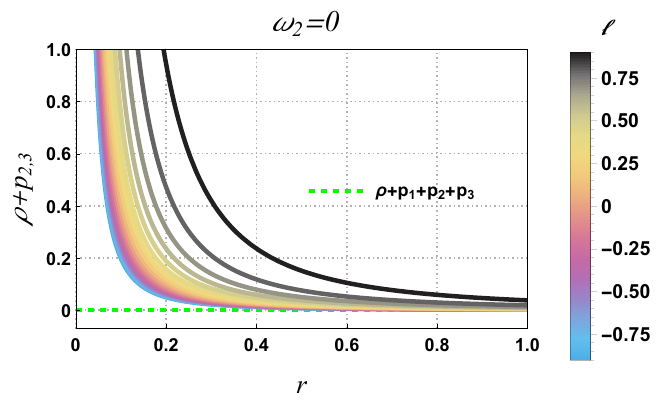} \\
        \footnotesize (a) [$K=0.095$] & \footnotesize (b) [$K=0.095$]
    \end{tabular}
    \caption{\footnotesize The variation of $\rho+\sum_i p_i$ (strong energy condition), $\rho+p_{2,3} $ (null energy condition), and $\rho-\mid p_{2,3}\mid $ (dominant energy condition) against $r$ for various values of $\ell$ with $w_2=0$.}
    \label{fig:model1_embeddings}
\end{figure}

Figure~\ref{fig:model1_embeddings} shows radial profiles of $\rho$, \(\rho+p_{2,3}\) and \(\rho+\sum_i p_i\) for \(K=0.095\), \(M=1\) and varying \(\ell\) with \(w_2=0\). Hence, the tangential NEC is algebraically equivalent to
\begin{equation}
K(1-\ell)\ge 0,
\end{equation}
the SEC is saturated (\(\rho+\sum_i p_i=0\)), and the DEC reduces to \(K(1-\ell)\ge0\) together with \(|w_2|\le1\) (here \(|w_2|=0\)). The observed decrease in the amplitude of \(\rho+p_{2,3}\) as \(\ell\) approaches unity follows directly from \(K(1-\ell)\): larger \(\ell\) reduces \(K(1-\ell)\) and pushes the tangential combination toward zero. The radial scaling for \(w_2=0\) is \(r^{-2}\), so the steep rise of the plotted combinations near the left axis corresponds to the core divergence implied by this exponent.

\begin{figure}[ht!]
    \centering
    \begin{tabular}{cc}
        \includegraphics[width=0.48\textwidth]{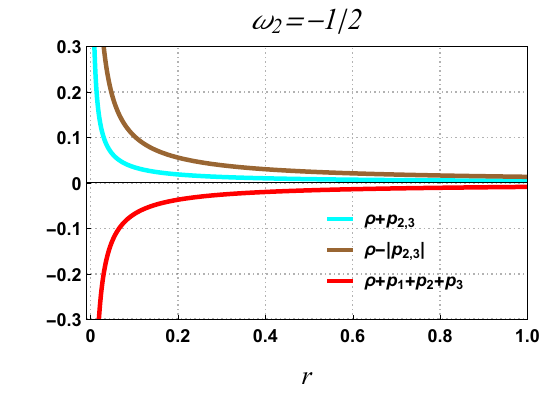} & 
        \includegraphics[width=0.48\textwidth]{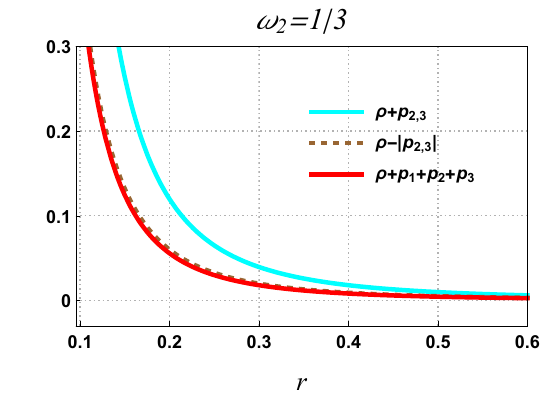} \\
        \footnotesize (a) [$K=0.095$, $\ell=0.1$]] & \footnotesize (b) [$K=0.095$, $\ell=0.1$]
    \end{tabular}
    \caption{\footnotesize The variation of $\rho+\sum_i p_i$ (strong energy condition), $\rho+p_{2,3} $ (null energy condition), and $\rho-\mid p_{2,3}\mid $ (dominant energy condition) against $r$ for $w_2=-1/2$ (left) and $w_2=1/3$ (right).}
    \label{fig:model2_embeddings}
\end{figure}

Figure~\ref{fig:model2_embeddings} presents the parameter-dependent behavior of \(K=0.095\) with two representative equations of state: \(w_2=-1/2\) (left panel) and \(w_2=1/3\) (right panel), using \(\ell=0.1\) in treatment. The controlling factor is evaluated to
\begin{equation}
K(2-\ell)\mid_{w_2=-\tfrac12}>0,\qquad K\bigl(\tfrac13-\ell\bigr)\mid_{w_2=\tfrac13}>0\ \text{(for }\ell=10^{-1}\text{)}.
\end{equation}
The energy-condition pattern follows from the algebraic sign rules:
\begin{itemize}
  \item For \(w_2=-1/2\): \(1+w_2=+1/2>0\) and \(K(2-\ell)>0\), so the tangential NEC \((1+w_2)K(2-\ell)\) is positive; \(|w_2|<1\) and \(K(2-\ell)>0\) imply that DEC holds pointwise; however, \(w_2<0\) with \(K(2-\ell)>0\) yields \(w_2 K(2-\ell)<0\) and therefore \(\rho+\sum_i p_i<0\), i.e., SEC violation across the plotted radii. The radial slopes are mild, consistent with the less singular exponents for this choice of \(w_2\).
  \item For \(w_2=1/3\): \(w_2>0\) and \(1+w_2>0\), so NEC, SEC and DEC reduce to the single algebraic requirement \(K\bigl(\tfrac13-\ell\bigr)\ge0\) (together with \(|w_2|\le1\) for DEC). With \(\ell=10^{-1}\) and \(K=0.095\) one obtains \(K\bigl(\tfrac13-\ell\bigr)>0\), hence all three conditions are satisfied pointwise; the radial decay is stronger and produces a sharper falloff at larger \(r\).
\end{itemize}
The two panels illustrate the algebraic pattern: \(K(1-2w_2-\ell)\) controls the sign of NEC and DEC combinations, while the product \(w_2K(1-2w_2-\ell)\) controls the SEC. 

\section{GEODESIC MOTIONS: NULL GEODESIC} \label{sec3}

This section examines the geodesic motion of test particles within the spacetime of an anisotropic fluid black hole in KR gravity. We focus on how the KR field parameter and the surrounding fluid density influence particle trajectories and light deflection. 
The study of geodesic motion is essential for understanding BH properties, as it provides insights into spacetime geometry and observable phenomena such as gravitational lensing and BH shadows \cite{Afrin:2021wlj,Kumar:2020,Ali:2024ssf,Ali:2025beh}. Analyzing the paths of particles and light is also crucial for interpreting high-energy astrophysical processes.
We will use the Lagrangian approach to investigate this motion \cite{Al-Badawi2026,ALBADAWI2025185,Al-Badawi2023,FA5,FA6}. 

 The Lagrangian density for the given metric is presented as the starting point for this analysis.
\begin{equation}
   \mathcal{L}=\frac{1}{2}\,\left[-f(r)\,\left(\frac{dt}{d\tau}\right)^2+\frac{1}{f(r)}\,\left(\frac{dr}{d\tau}\right)^2+r^2\left(\frac{d\phi}{d\tau}\right)^2\right],\label{mm1}
\end{equation}
where $\tau$ represents an affine parameter and the geodesic motion in the equatorial plane  $\theta=\pi/2$.\\
Since the chosen spacetime is static and spherically symmetric, it admits two Killing vector fields:  the energy $\mathrm{E}$ and the angular momentum $\mathrm{L}$ of test particles, which are given by
\begin{equation}
   \mathrm{E}=f(r)\,\left(\frac{dt}{d\tau}\right)\quad,\quad \mathrm{L}=r^2\,\frac{d\phi}{d\tau}.\label{mm2}
\end{equation}
After substituting  Eq. (\ref{mm2}) into Eq. (\ref{mm1}), we obtain the equation of motion associated with the radial coordinate $r$.
\begin{equation}
   \left(\frac{dr}{d\tau}\right)^2+V_\text{eff}(r)=\mathrm{E}^2\label{mm3}
\end{equation}
which is equivalent to the one-dimensional equation of motion of a unit mass particle having energy $\mathrm{E}^2$ and the potential  $V_\text{eff}(r)$. The effective potential governing the dynamics of test particles around the BH is given by 
\begin{equation}
   V_\text{eff}(r)=\left(\varepsilon+\frac{\mathrm{L}^2}{r^2}\right)\,\left(\frac{1}{1-\ell}-\frac{2 M}{r}-\frac{K}{1-\ell}\,r^{\frac{2 w_2}{\ell-1}}\right).\label{mm4}
\end{equation}
Here $\varepsilon=0$ for null geodesics and $1$ for time-like geodesics.
\subsection{Null Geodesic}
The study of null geodesics is essential for understanding black hole physics and their observable features, like effective force, photon sphere and shadow. The effective potential is a key tool for this analysis, as it describes how light behaves in the curved spacetime near a black hole.
In the null geodesic case, $\varepsilon=0$, the effective potential from Eq. (\ref{mm4}) becomes:
\begin{equation}
   V_\text{eff}(r)=\frac{\mathrm{L}^2}{r^2}\,\left(\frac{1}{1-\ell}-\frac{2 M}{r}-\frac{K}{1-\ell}\,r^{\frac{2 w_2}{\ell-1}}\right).\label{cc1}
\end{equation}

Here, we study the dynamics of photons in a gravitational field and show how various parameters affect their effective radial force near a BH. Using the effective potential  given in Eq. (\ref{cc1}), we can determine the effective radial force on as,
\begin{equation}
   \mathrm{F}_\text{ph}=-\frac{1}{2}\,\frac{dV_\text{eff}}{dr}=\frac{\mathrm{L}^2}{r^3}\,\left(\frac{1}{1-\ell}-\frac{3\,M}{r}-\frac{K(1+w_2)}{(1-\ell)r^{2w_2}}\right).\label{cc2}
\end{equation}

We see that the effective radial force experienced by the photon particles is influenced by the KR parameter $\ell$, the fluid density parameter $K$, the conserved angular momentum $\mathrm{L}$ and the BH mass $M$. In the limit where $K=0$, the above result (\ref{cc2}) reduces to that of the Schwarzschild BH solution in KR gravity, which further simplifies to the standard Schwarzschild BH result when $\ell=0$. The effective radial force for three distinct values of the equation-of-state parameter is shown in Table \ref{tab:5}.

\begin{table}[ht!]
\centering
\begin{tabular}{|l|l|l|}
\hline
\textbf{Case} & \textbf{Effective Radial Force }  \\
\hline
\( w_2 = 0 \) (Dust) & 
$\frac{\mathrm{L}^2}{r^3}\,\left(\frac{1}{1-\ell}-\frac{3\,M}{r}-\frac{K}{1-\ell}\right)$ 
\\
\hline

\(w_2 = 1/3 \) (Radiation)& 
\( \frac{\mathrm{L}^2}{r^3}\,\left(\frac{1}{1-\ell}-\frac{3\,M}{r}+\frac{K(3\ell-4)}{3(1-\ell)^2}r^{2/3(\ell-1)}\right)\)
 \\
\hline

\( w_2 = -1/2\) (Dark Energy-like)& 
\( \frac{\mathrm{L}^2}{r^3}\,\left(\frac{1}{1-\ell}-\frac{3\,M}{r}+\frac{K(2\ell-1)}{2(1-\ell)^2}r^{1/(1-\ell)}\right)\)  
.

\\
\hline
\end{tabular}
\caption{Effective radial force under values of the equation-of-state parameter.}
\label{tab:5}
\end{table}

\subsection{Photon sphere and  BH shadow}

This subsection analyzes the combined impact of KR gravity and an anisotropic fluid on key features of photon dynamics: the photon sphere and shadow radius.

Circular null geodesics require the conditions $\dot{r}=0$ and $\ddot{r}=0$. This leads to the following two relationships:
\begin{equation}
    \mathrm{E}^2=V_\text{eff}(r)=\frac{\mathrm{L}^2}{r^2}\,f(r)\quad,\quad \frac{dV_\text{eff}(r)}{dr}=0.\label{cc4} 
\end{equation}

The first relation in Eq. (\ref{cc4}) gives us the critical impact parameter for photons. The second relation $\frac{dV_\text{eff}(r)}{dr}=0$ gives us the photon sphere radius $r=r_\text{ph}$ satisfying the following equation:
\begin{equation}
    2\,f(r)=r\,f'(r)\Rightarrow 6(1-\ell)^2M-2r\left(    1-\ell+K(\ell-1-w_2)r^{2w_2/(\ell-1)}\right)
 =0.\label{eps1}
\end{equation}

When the parameters are set to $K=0$ and $\ell=0$, Eq. (\ref{eps1}) simplifies to $3M$. The solution of Eq. (\ref{eps1}) depends on the choice of the equation-of-state parameter $w_2$. Analytically, we obtain the photon sphere radius for three distinct values of the equation-of-state parameter as follows:\\
\noindent\textbf{Case I: \( w_2 = 0 \) (Dust)} The photon sphere equation (\ref{eps1}) becomes 
\begin{equation}
   (1-K)r-3(1-\ell)M=0 \quad \Rightarrow \quad r_{ph} =\frac{3\,M(1-\ell)}{1-K}. \label{phw0}
\end{equation}
\noindent\textbf{Case II: \( w_2 = \frac{1}{3} \) (Radiation)} Equation (\ref{eps1}) becomes 
\begin{equation} 
\left(1-\ell +K(\ell-\frac{4}{3})r^{2/3(\ell-1)}\right)r-3(1-\ell)^2M=0 . \label{photon133}\end{equation} 
For $0<K<1$, the photon sphere equation (\ref{photon133}) reduces to a nonlinear algebraic equation with fractional powers. By introducing the variable $x=r^{1/3}$, the equation can be recast into a mixed-power polynomial form as \begin{equation}
    (1-\ell)x^3+K(\ell-\frac{4}{3})x^{2\ell+1}-3(1-\ell)^2M=0.
\end{equation} Although no closed-form solution exists for generic $\ell$, the equation admits a unique positive real root. Exact analytical solutions for the photon sphere radius are challenging. Thus, we attempted a numerical solution to determine the photon sphere radius.\\
\noindent\textbf{Case III: \( w_2 = -\frac{1}{2} \) (Dark Energy-like)} Equation (\ref{eps1}) becomes
\begin{equation} 
\left(1-\ell +K(\ell-\frac{1}{2})r^{1/(1-\ell)}\right)r-3(1-\ell)^2M=0 . \label{phton12}\end{equation} 
 Approximate solution of Eq. (\ref{phton12}) is given by (see appendix)
\begin{equation}
    r(K) \approx 3(1-\ell)M 
\;-\; \frac{\ell-\tfrac12}{1-\ell}\,
\bigl[3(1-\ell)M\bigr]^{\frac{2-\ell}{1-\ell}} K
\;+\; \mathcal{O}(K^2) \label{photon14}
\end{equation}

\begin{figure}
\begin{center}
\includegraphics[scale=0.4]{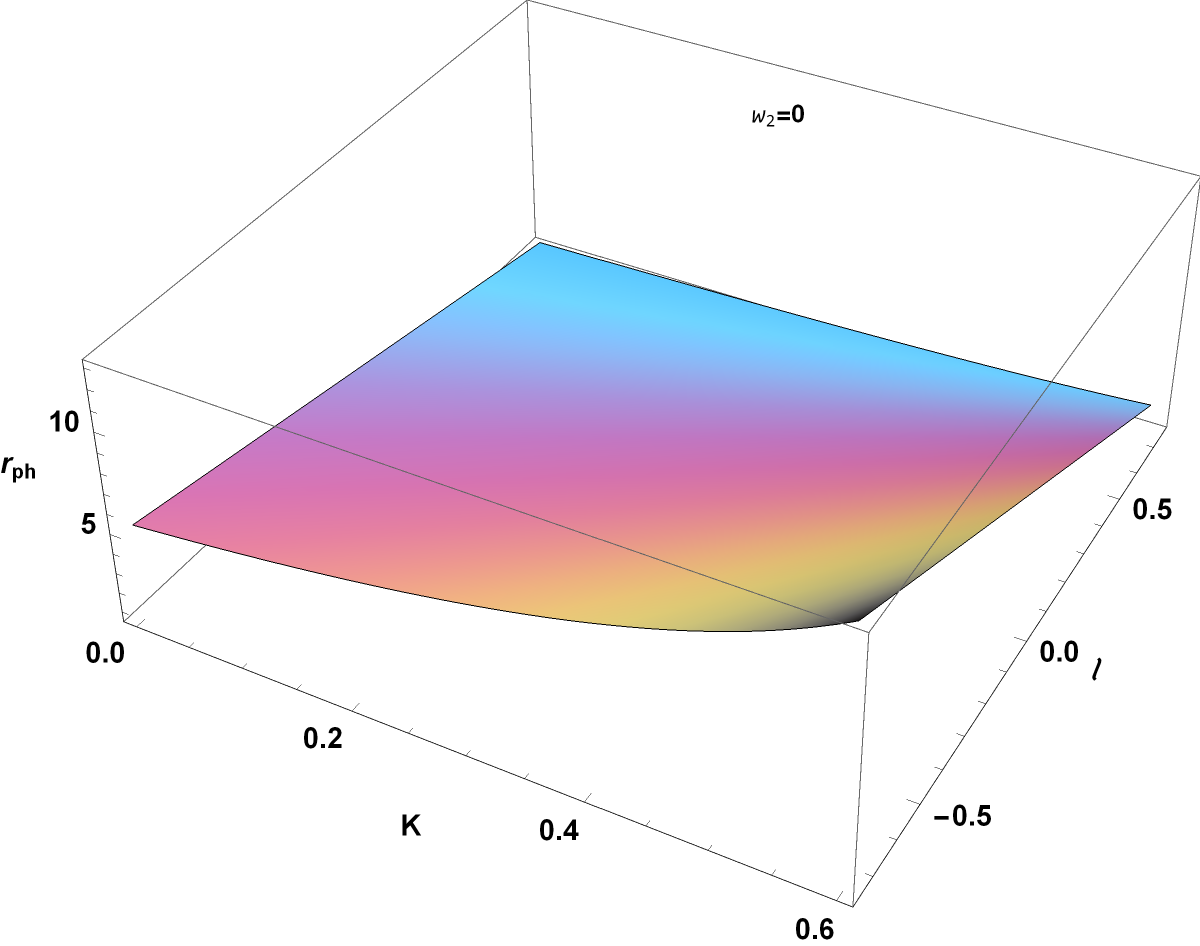}
\includegraphics[scale=0.4]{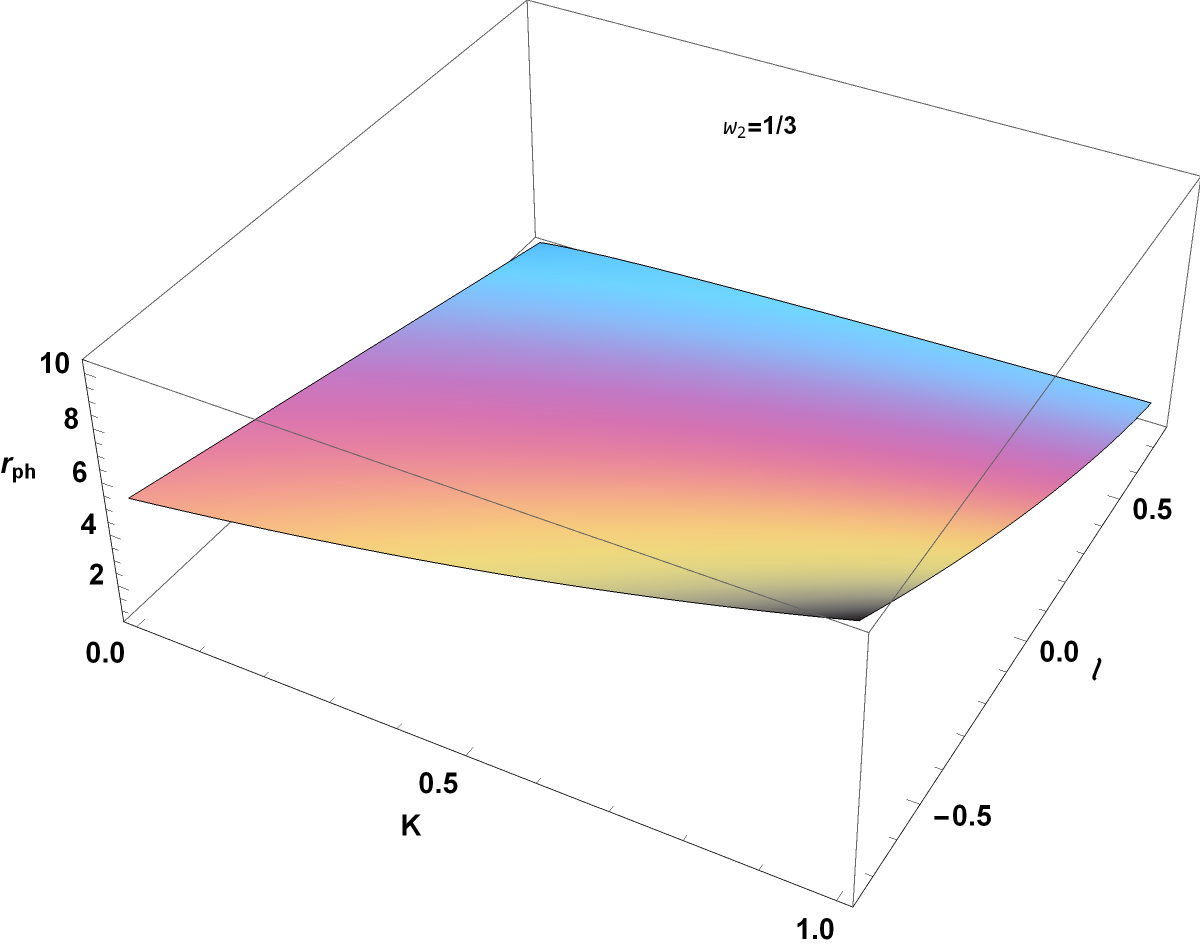}\includegraphics[scale=0.4]{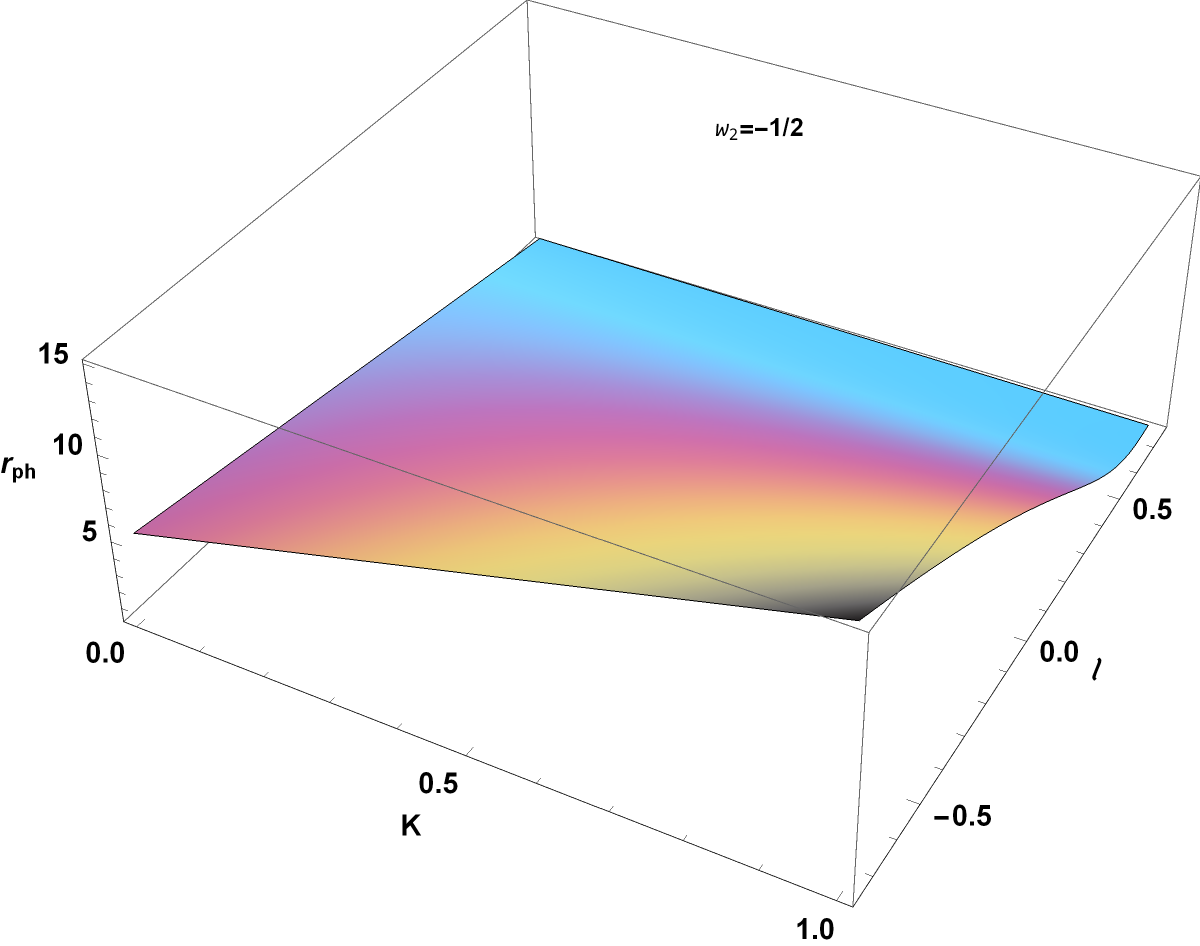}
\end{center}
\caption{\footnotesize Plot of the photon sphere $r_{ph}$ vs $K$ and  $\ell$ and for certain values of $w_2$, dust (Top), Radiation  (middle), and dark energy-like (bottom). Here, $M=1$.}\label{photon1}
\end{figure}

Figure \ref{photon1} shows three-dimensional visualizations of the photon sphere radius for three distinct
values of the equation-of-state parameter $w_2$ as a function of the combined values of $K$ and $\ell$.  Our analysis shows that raising both $K$ and $\ell$ expands the photon sphere $r_{ph}$.

Next, we continue computing the black hole shadow, the dark image formed by photons trapped near unstable circular orbits at the photon sphere. Its measurable radius $R_s$ is determined by the critical impact parameter $b_c$, which depends on the spacetime geometry. 
The BH shadow, as observed by a static observer at radial position $r_O$, has an apparent radius given by \cite{Volker2022}
\begin{equation}
    R_{\rm sh}=r_{\rm ph}\sqrt{\frac{f(r_{O})}{f(r_{\rm ph})}}=r_{\rm ph}\sqrt{\frac{ \frac{1}{1-\ell}-\frac{2 M}{r_O}-\frac{K}{1-\ell}\,r_O^{\frac{2 w_2}{\ell-1}}}{ \frac{1}{1-\ell}-\frac{2 M}{r_{\rm ph}}-\frac{K}{1-\ell}\,r_{\rm ph}^{\frac{2 w_2}{\ell-1}}}}.\label{shadow}
\end{equation}
 Analytically, the shadow radius for three distinct values of the equation-of-state parameter is :\\
\noindent\textbf{Case I: \( w_2 = 0 \) (Dust)}
\begin{equation}
     R_{\rm sh}=r_{\rm ph}\sqrt{\frac{ \frac{1}{1-\ell}-\frac{2 M}{r_O}-\frac{K}{1-\ell}}{ \frac{1}{1-\ell}-\frac{2 M}{r_{\rm ph}}-\frac{K}{1-\ell}}}
\end{equation}
For a distant observer ($r_O \to \infty$), the shadow radius simplifies to
\begin{equation}
    R_{\rm sh}=r_{\rm ph}\sqrt{\frac{\frac{1-K}{1-\ell} }{ \frac{1}{1-\ell}-\frac{2 M}{r_{\rm ph}}-\frac{K}{1-\ell}}}\,.\label{shadow9}
\end{equation}
\noindent\textbf{Case II: \( w_2 = \frac{1}{3} \) (Radiation)}
\begin{equation}
    R_{\rm sh}=r_{\rm ph}\sqrt{\frac{ \frac{1}{1-\ell}-\frac{2 M}{r_O}-\frac{K}{1-\ell}\,r_O^{\frac{2}{3(\ell-1)}}}{ \frac{1}{1-\ell}-\frac{2 M}{r_{\rm ph}}-\frac{K}{1-\ell}\,r_{\rm ph}^{\frac{2 }{3(\ell-1)}}}}.
\end{equation}
For a distant observer ($r_O \to \infty$) and $-1<\ell<1$, the shadow radius simplifies to
\begin{equation}
    R_{\rm sh}=\,r_{\rm ph}\sqrt{\frac{\frac{1}{1-\ell} }{\frac{1}{1-\ell}-\frac{2 M}{r_{\rm ph}}-\frac{K}{1-\ell}\,r_{\rm ph}^{\frac{2 }{3(\ell-1)}}}}.\label{shadow91}
\end{equation}
\\
\noindent\textbf{Case III: \( w_2 = -\frac{1}{2} \) (Dark Energy-like)}
\begin{equation}
    R_{\rm sh}=r_{\rm ph}\sqrt{\frac{ \frac{1}{1-\ell}-\frac{2 M}{r_O}-\frac{K}{1-\ell}\,r_O^{\frac{1}{1-\ell}}}{ \frac{1}{1-\ell}-\frac{2 M}{r_{\rm ph}}-\frac{K}{1-\ell}\,r_{\rm ph}^{\frac{1}{1-\ell}}}}.\label{shadow09}
\end{equation}
For a distant observer ($r_O \to \infty$), the shadow radius
    $R_{\rm sh}=\text{Undefined}$.

\begin{figure}
\begin{center}
\includegraphics[scale=0.4]{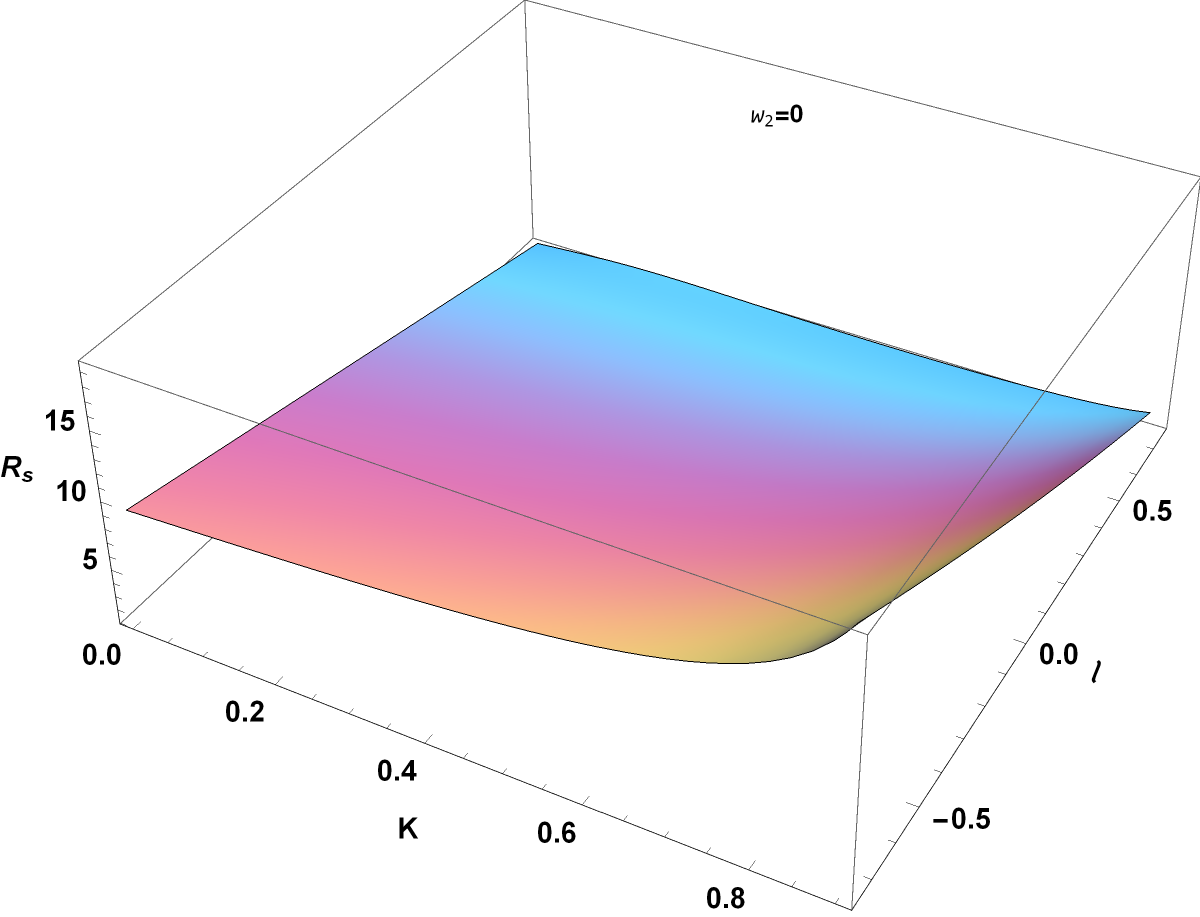}
\includegraphics[scale=0.4]{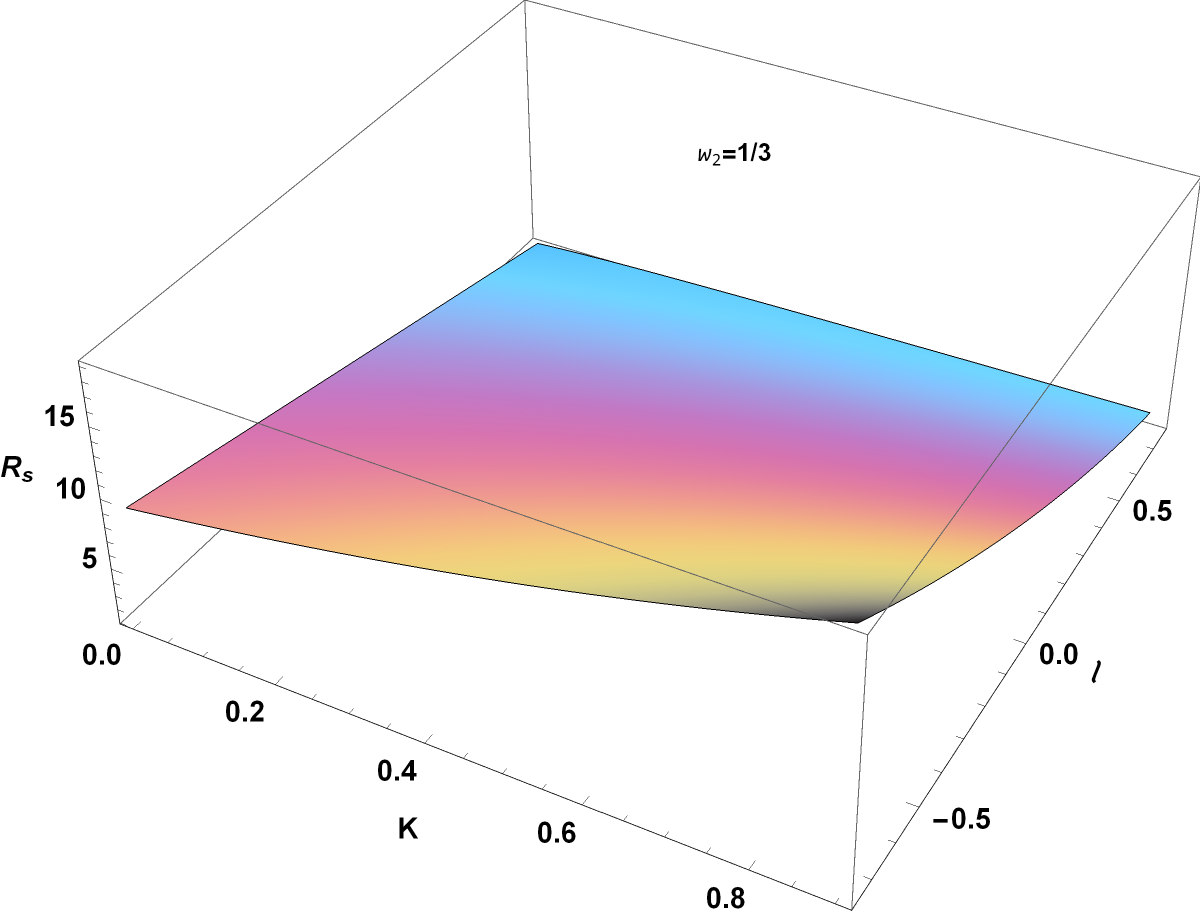}
\end{center}
\caption{\footnotesize Plot of the shadow radius $R_{s}$ vs $K$ and $\ell$ and for certain values of $w_2$, Dust (left) and  Radiation  (right). Here, $M=1$.}\label{shadow1}
\end{figure}

Figure \ref{shadow1} shows three-dimensional visualizations of the shadow radius for two distinct
values of the equation-of-state parameter $w_2$ as a function of the combined values of $K$ and $\ell$.  The figure shows that the shadow radius increases with both parameters $K$ and $\ell$.

\section{Weak deflection angle}\label{sec:weak_deflection}


Gravitational lensing arises from the bending of light by the gravitational field of massive objects such as planets, black holes, or dark matter, a phenomenon predicted by Einstein’s general relativity in the weak-field regime. In particular, weak gravitational deflection plays a central role in observational astrophysics, as it is widely used to trace dark matter filaments and to probe the large-scale structure of the Universe.
Among the available techniques to compute the weak deflection angle, a powerful geometric approach was developed by Gibbons and Werner, based on the Gauss–Bonnet theorem applied to the optical metric \cite{gibbons_applications_2008,werner_gravitational_2012}. Within this framework, the bending of light can be interpreted as a global geometric (partly topological) effect, and the deflection angle is obtained by integrating the Gaussian curvature of the optical manifold outside the photon trajectory. Thanks to its geometric formulation, this method has been widely used in many lensing scenarios \cite{jusufi_light_2017,crisnejo_weak_2018,ovgun_gravitational_2018,jusufi_effect_2018,ovgun_weak_2019,javed_effect_2019,de_leon_weak_2019,ovgun_weak_2019-1,li_equivalence_2020,javed_effect_2020,javed_weak_2020,javed_weak_2020-1}).

In this subsection, we pursue the approach in Refs.  \cite{Ishihara2016,Ishihara2017}, which introduces a method to calculate the light bending angle for non-asymptotically flat spacetimes.

For null geodesics ($ds^2=0$), the optical metric is defined by
\begin{equation}
dt^2 = \bar{g}_{ij}dx^{i} dx^{j},
    \label{eq:optical_metric}
\end{equation}
which yields
\begin{equation}
d\sigma^2 = \frac{dr^2}{f(r)^2} + \frac{r^2}{f(r)}d\phi^2.
    \label{eq:dsigma}
\end{equation}
This two-dimensional Riemannian geometry fully encodes light propagation. In such a case, the Gaussian curvature associated with the optical metric is obtained as
\begin{equation}
\mathcal{K} = - \frac{1}{r}{f(r)\sqrt{f(r)}}\,\frac{d}{dr}\left[f(r) \frac{d}{dr}\left[\frac{r}{\sqrt{f(r)}}\right]\right],
    \label{eq:K}
\end{equation}
which can be approximated as
\begin{equation}
\mathcal{K} = -\frac{2M}{r^3} - \frac{K}{2(1-\ell)}\nu(\nu-1)r^{\nu-2} + \mathcal{O}(M^2,K^2,MK),
    \label{eq:K_1}
\end{equation}
where we have defined
\begin{equation}
\nu \equiv \frac{2 w_2}{\ell-1}.
    \label{eq:nu}
\end{equation}
The expression \eqref{eq:K_1} is essentially what is needed to apply the Gauss-Bonnet theorem used in Refs. \cite{Ishihara2016,Ishihara2017}. We consider the quadrilateral domain $\mathcal{D}$ bounded by the photon trajectory $\gamma$, a radial geodesic from the source $S$ at $r_S$, and a radial geodesic from the observer $O$ at $r_O$, a circular arc $C_R$ of radius $R$, which will be shrunk away. The Gauss-Bonnet theorem then gives
\begin{equation}
\iint_\mathcal{D} \mathcal{K}\,d\mathcal{S} + \int_\gamma \kappa_g dl + \sum_i \theta_i = 2\pi,
    \label{eq:GB_0}
\end{equation}
where $\kappa_g$ is the geodesic curvature of the light ray boundary curve with respect to the optical metric, and along the photon trajectory, we have $\kappa_g=0$. After careful evaluation, the finite-distance deflection angle is \cite{Ishihara2017}
\begin{equation}
\hat\alpha = - \iint_\mathcal{D} \mathcal{K}\, d\mathcal{S} + \Psi_O + \Psi_S - \pi, 
    \label{eq:alpha_0}
\end{equation}
in which $\Psi_O$ is the angle between the photon trajectory and the radial direction at the observer, and $\Psi_S$ is the corresponding angle at the source.

The surface element of the optical metric is
\begin{equation}
d\mathcal{S} = \sqrt{\mathrm{det}[\bar{g}]}\, dr \,d\phi = \frac{r}{f(r)^{3/2}}\,dr \,d\phi.
    \label{eq:dS}
\end{equation}
To leading-order, the photon trajectory is approximated by
\begin{equation}
r(\phi) \simeq \frac{b}{\sin\phi}, 
    \label{eq:rphi}
\end{equation}
where the impact parameter $b\equiv \mathrm{L}/\mathrm{E}$, is defined at the point of closest approach as
\begin{equation}
b^2 = \frac{r_0^2}{f(r_0)},
    \label{eq:b}
\end{equation}
which remains well-defined without asymptotic flatness. The radial integration is therefore bounded by
\begin{equation}
r(\phi)\leq r \leq r_O\quad ({\text{observer side}}),\qquad r(\phi)\leq r\leq r_S\quad ({\text{source side}}).
    \label{eq:geodesics_bounded}
\end{equation}
Now, splitting the domain at the point of closest approach, the curvature contribution becomes
\begin{equation}
\hat{\alpha}_\mathcal{K} = - \int_{0}^{\phi_O}\int_{b/\sin\phi}^{r_O} \mathcal{K}\,d\mathcal{S} - \int_{\phi_O}^{\pi}\int_{b/\sin\phi}^{r_S} \mathcal{K}\,d\mathcal{S}.
    \label{eq:alpha_K}
\end{equation}
We now evaluate this explicitly. 

Using the expression in Eq. \eqref{eq:K_1}, one gets the contribution of the Schwarzschild part to the deflection angle as
\begin{equation}
\hat{\alpha}_M = \int\frac{2 M}{r^3}\,\frac{r}{f_0^{3/2}}\,dr\,d\phi,
    \label{eq:alpha_M}
\end{equation}
where $f_0 = 1/(1-\ell)$. This yields
\begin{equation}
\hat{\alpha}_M = \frac{2 M}{f_0^{3/2}}\left[\int_0^{\phi_O}\left(\frac{1}{b/\sin\phi}-\frac{1}{r_O}\right)d\phi + \int_{\Phi_O}^\pi\left(\frac{1}{b/\sin\phi}-\frac{1}{r_S}\right)d\phi\right].
    \label{eq:alpha_M1}
\end{equation}
This integral can be calculated directly, yielding
\begin{equation}
\hat{\alpha}_M = \frac{2M}{b}f_0^{-3/2}\left(\sin\phi_O + \sin\phi_S\right)-2 M f_0^{3/2}\left(\frac{\phi_O}{r_O}+\frac{\pi+\phi_O}{r_S}\right).
    \label{eq:alpha_M2}
\end{equation}
Now the local angles at the source and observer satisfy the relations
\begin{equation}
\sin\Psi_O = \frac{b\sqrt{f(r_O)}}{r_O},\qquad \sin\Psi_S = \frac{b\sqrt{f(r_S)}}{r_S}.
    \label{eq:PsiOPsiS}
\end{equation}
Geometrically this means
\begin{equation}
\phi_O = \pi - \Psi_O,\qquad  \phi_S = \Psi_S.
    \label{eq:phiOphiS}
\end{equation}
Accordingly, the total deflection angle is given by
\begin{equation}
\hat\alpha = \hat\alpha_M + \hat\alpha_K  + \Psi_O + \Psi_S - \pi.
    \label{eq:alpha_total0}
\end{equation}
Now to calculate the contribution of the dark matter, i.e., the $K$-term, we use the second term of Eq. \eqref{eq:K_1}, which gives
\begin{equation}
\hat\alpha_K = \frac{f_0}{2}K(\nu-1)\left\{\int_0^{\phi_0}\left[\left(\frac{b}{\sin\phi}\right)^\nu-r_O^\nu\right] d\phi + \int_{\phi_0}^\pi\left[\left(\frac{b}{\sin\phi}\right)^\nu-r_S^\nu\right] d\phi\right\}.
    \label{eq:alpha_K0}
\end{equation}
For the case of a universe filled with dust (i.e. $w_2=0$ or $\nu=0$), this integral provides, after manipulations
\begin{equation}
\hat\alpha_{K}^{\mathrm{dust}} = -\frac{\pi}{2} K f_0.
    \label{eq:alpha_K_dust}
\end{equation}
For the radiation case (i.e. $w_2 = 1/3$ or $\nu = 2f_0/3$), the integral in Eq. \eqref{eq:alpha_K0}, provides
\begin{equation}
\hat\alpha_K^{\mathrm{rad}} = \frac{f_0}{2}K(\nu-1)\left[b^\nu\sqrt{\pi}\,\frac{\displaystyle\Gamma\left(\frac{1-\nu}{2}\right)}{\displaystyle\Gamma\left(\frac{2-\nu}{2}\right)} - \Bigl(\phi_O r_O^\nu+(\pi-\phi_O)r_S^\nu\Bigr)\right],
    \label{eq:alpha_K_radiation}
\end{equation}
to obtain which, we have used the identity
\begin{equation}
\int_0^\pi\sin^{-\nu}\phi\,d\phi = \sqrt{\pi}\,\frac{\displaystyle\Gamma\left(\frac{1-\nu}{2}\right)}{\displaystyle\Gamma\left(\frac{2-\nu}{2}\right)}.
    \label{eq:I_nu}
\end{equation}
For a universe dominated by DE-like matter, characterized by $w_2 = -1/2$ or equivalently $\nu = f_0$, the same procedure applied to the master integral in Eq.~\eqref{eq:alpha_K0} leads to an analytic expression identical to that obtained in Eq.~\eqref{eq:alpha_K_radiation}. The distinction arises at the numerical level, where the appropriate value of $\nu$ must be implemented when generating the corresponding profiles.

\begin{figure}[t]
    \centering
    \includegraphics[width=5.3 cm]{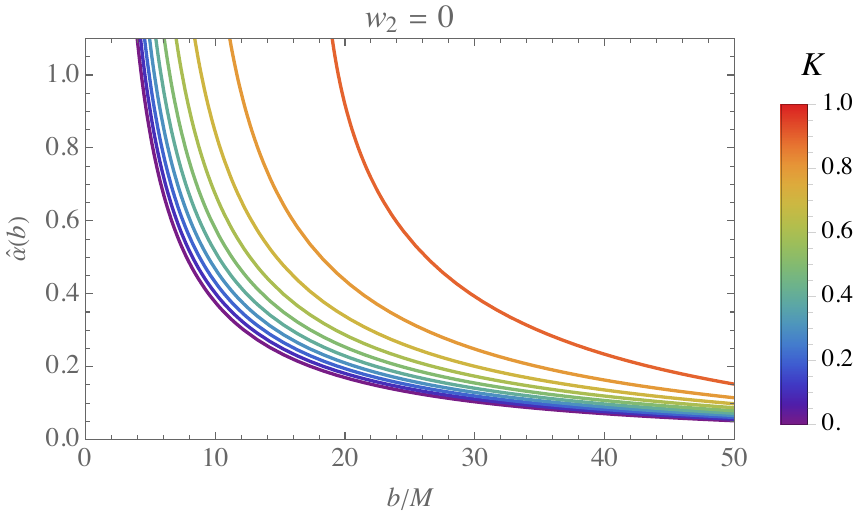} (a)
     \includegraphics[width=5.3 cm]{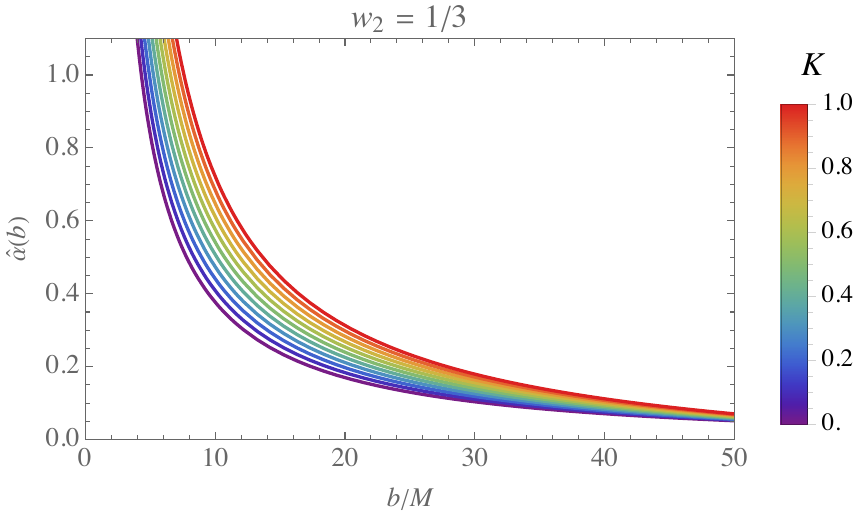} (b) 
     \includegraphics[width=5.3 cm]{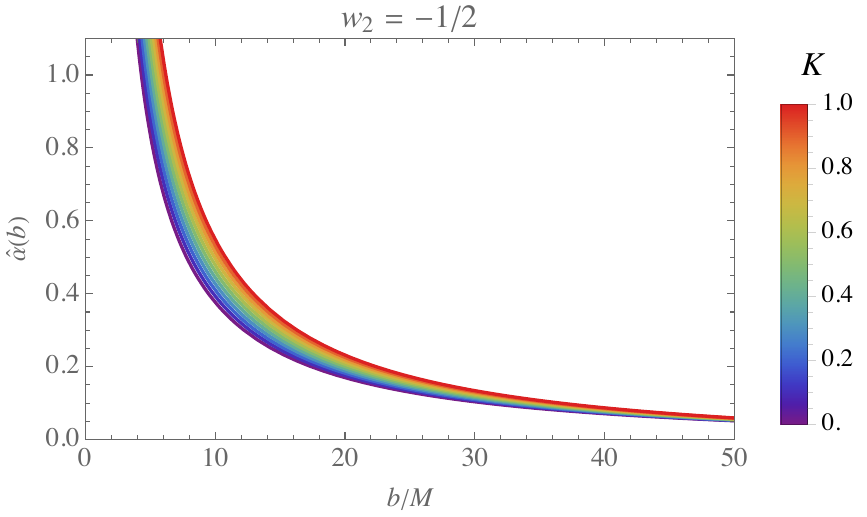} (c)
     \includegraphics[width=5.3 cm]{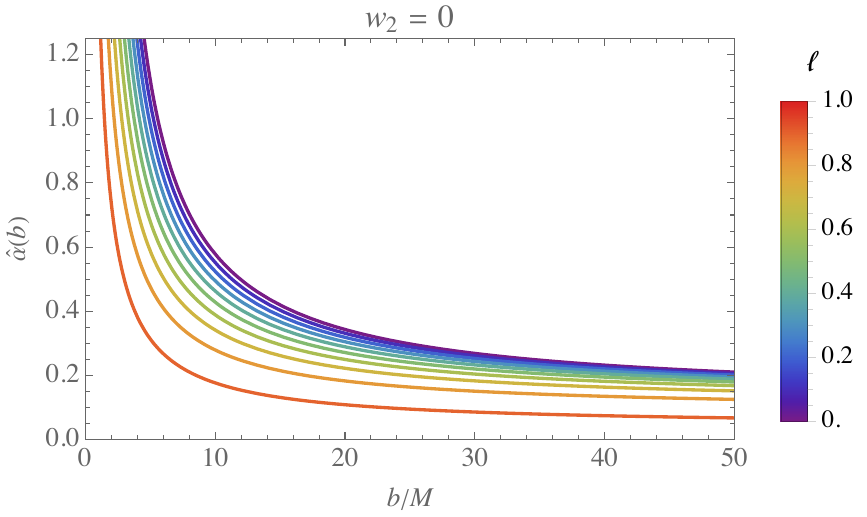} (d)
     \includegraphics[width=5.3 cm]{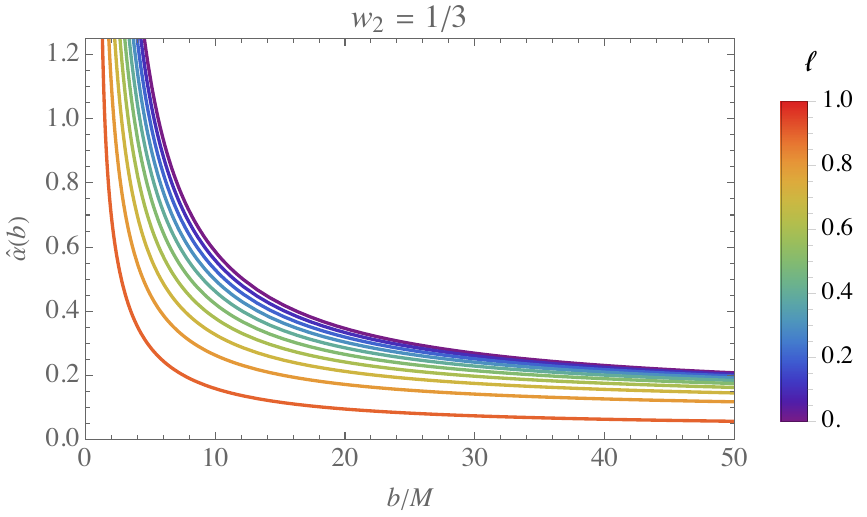} (e)
     \includegraphics[width=5.3 cm]{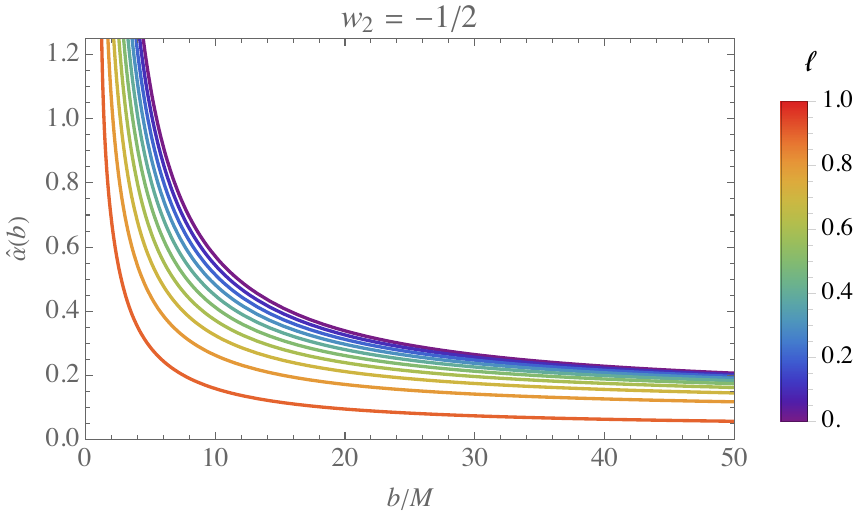} (f)
    \caption{The $b$-profiles of the weak deflection angle $\hat{\alpha}(b)$ for the three values of $w_2$, evaluated at $r_O = r_S = 100 M$. Panels (a–c) correspond to fixed $\ell = 0.1$ and varying $K$, while panels (d–f) show fixed $K = 0.1$ and varying $\ell$.}
    \label{fig:alpha}
\end{figure}

In Fig.~\ref{fig:alpha}, we display several $b$-profiles of the weak deflection angle for the three considered values of $w_2$, allowing a direct comparison of the lensing behaviour across the different cosmological backgrounds.
From the diagrams, it is clear that, in all configurations, the deflection angle decreases monotonically with increasing impact parameter $b$, while larger values $K$ enhance the overall bending, consistently reflecting a stronger effective gravitational field. For dust ($w_2=0$), the curves show only a moderate spread, indicating a weak sensitivity of the deflection angle to both $K$ and the KR parameter $\ell$. This sensitivity becomes more pronounced for radiation ($w_2=1/3$), where the profiles are steeper at small $b$ and the separation between curves increases, signaling a stronger interplay between geometry, matter content, and the $\ell$ spectrum. The DE-like case ($w_2=-1/2$) exhibits the largest deviations, with both $K$ and $\ell$ producing a significant enhancement of the deflection angle over a wide range of impact parameters, particularly in the strong-lensing regime. Thus, while the qualitative behaviour of $\hat{\alpha}(b)$ is universal, its magnitude and parametric sensitivity grow as one moves from dust to radiation and finally to DE-dominated backgrounds.

\section{Gravitational lensing in SDL}\label{sec:strongdeflection}
In this section, we investigate strong gravitational lensing by an anisotropic fluid black hole within the Kalb-Ramond gravity framework. Specifically, we analyse the trajectory of light rays in the equatorial plane and examine how the black hole parameters influence the lensing observables in SDL, where the closest approach distance $r_0$ approaches the photon sphere radius $r_{\text{ph}}$. In this regime, the deflection angle increases monotonically, exceeding $2\pi$ radians and diverging logarithmically as $r_0 \to r_{\text{ph}}$ \cite{Bozza:2002zj}.

\begin{figure}[t]
    \centering
    \includegraphics[width=8.3 cm]{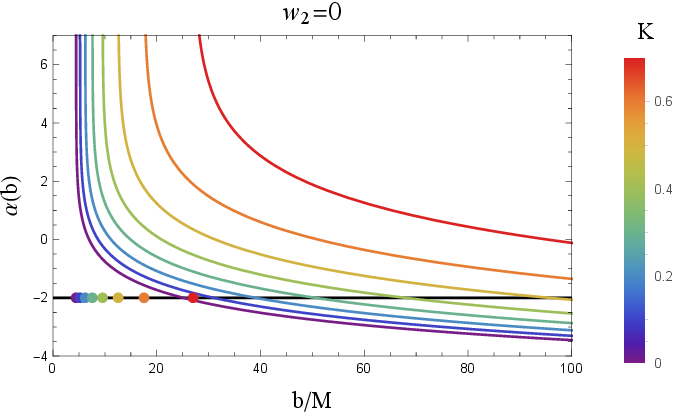} (a)
     \includegraphics[width=8.3 cm]{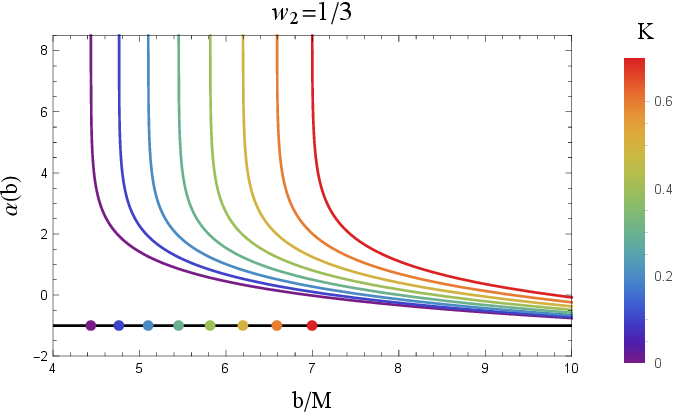} (b) 
     \includegraphics[width=8.3 cm]{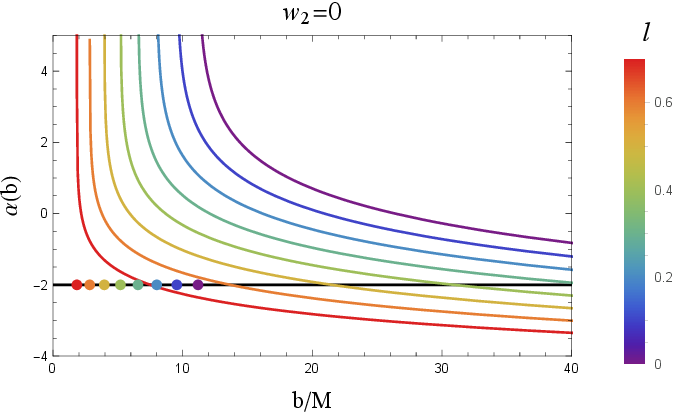} (c)
     \includegraphics[width=8.3 cm]{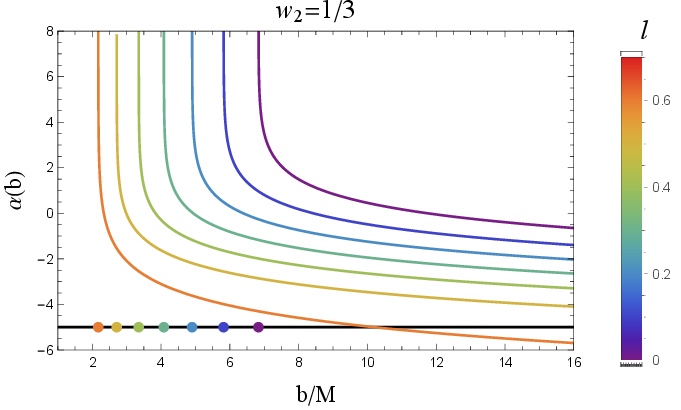} (d)
    \caption{The deflection angle as a function of $b$ for two values of $w_2$. Panels (a–b) correspond to fixed $\ell = 0.1$ and varying $K$, while panels (c–d) show fixed $K = 0.4$ and varying $\ell$. Here, the dots on the black solid line correspond to the values of the critical impact parameter where $\alpha(b)$ diverges.}
    \label{fig:alphastrong}
\end{figure}

For a photon propagating on the equatorial plane of a static and spherically symmetric spacetime, the bending of light is characterized by the deflection angle $\alpha_D(r_0)$, defined as the angle between the asymptotic incoming and outgoing directions. As a function of the closest approach distance $r_0$, it is given by \cite{Virbhadra:1998dy, Bozza:2002zj, Islam:2021ful, Islam:2021dyk,Vachher:2025jsq}:

\begin{equation}\label{bending2}
\alpha_{D}(r_0) = I(r_0) - \pi,
\end{equation}

where $I(r_0)$ represents the total azimuthal angle traversed by the photon from its point of closest approach to infinity.
Using the geodesic equations \eqref{mm2} and \eqref{mm3}, we can find an explicit expression for the integral $I(r_0)$ in terms of metric coefficients. 
Since the integral cannot be solved explicitly, the integral is expanded near the unstable photon sphere radius \cite{Virbhadra:1999nm,Bozza:2002zj} by defining a new variable $z=1-r_0/r$ in SDL \cite{Islam:2022ybr,Vachher:2024ezs}. The analytical expression of the deflection angle for spacetime (\ref{eq:metric}) as a function of the impact parameter ($b\approx\theta D_{OL}$) is given by \cite{Bozza:2002zj,Kumar:2020sag,Islam:2020xmy}
\begin{eqnarray}\label{def4}
\alpha_{D}(b) &=& \bar{a} \log\left(\frac{b}{b_c} -1\right) + \bar{b} + \mathcal{O}(b-b_c),  
\end{eqnarray}  
where $\bar{a}$, $\bar{b}$ are the strong lensing coefficients. Detailed calculations can be found in \cite{Bozza:2002zj,Kumar:2020sag,Islam:2020xmy}. Figure \ref{fig:alphastrong} shows the deflection angle for an anisotropic fluid black hole in KR gravity, with $w_2=0$ (Dust) and $w_2=1/3$ (Radiation), for different values of the parameters $\ell$ and $K$. For dust, we observe that the deflection angle diverges at larger values of the critical impact parameter than in the radiation case. In case of $w_2=-1/2$ (Dark energy like), however, the critical impact parameter is not defined, as indicated in Eq.~\eqref{shadow09}. As a result, strong deflection cannot occur from the perspective of a distant observer.

The theoretical framework for analyzing strong gravitational lensing in the vicinity of a black hole is completed by the lens equation, which relates the angular position of the source to the apparent positions of the resulting relativistic images. For a scenario where both the observer and the source are situated in an asymptotically flat spacetime region far from the lensing black hole, and are nearly perfectly aligned with it, the lens equation can be approximated as \cite{Bozza:2002zj, Bozza:2008ev}:

\begin{eqnarray}\label{lenseq}
\beta = \theta - \frac{D_{LS}}{D_{OL}+D_{LS}} \Delta\alpha_n,
\end{eqnarray}

where $\beta$ and $\theta$ denote the angular positions of the source and the image, respectively, measured from the optical axis (the line connecting the observer to the lens). The quantity $\Delta \alpha_n = \alpha(\theta) - 2n\pi$ represents the offset of the deflection angle from the $2n\pi$ multiple required for the photon to loop around the black hole $n$ times. In SDL, we have $0 < \Delta \alpha_n \ll 1$. Here, $D_{\mathrm{LS}}$ is the distance from the lens to the source plane, and $D_{\mathrm{OL}}$ is the distance from the observer to the lens, with the observer-source distance approximated as $D_{\mathrm{OS}} \approx D_{\mathrm{OL}} + D_{\mathrm{LS}}$. 

Following Bozza \cite{Bozza:2002zj,Kumar:2025bim}, we define three characteristic observables in SDL as 
\begin{eqnarray}
\theta_\infty &=& \frac{b_c}{D_{OL}},\label{theta}\\
s &=& \theta_1-\theta_\infty \approx \theta_\infty ~\text{exp}\left({\frac{\bar{b}}{\bar{a}}-\frac{2\pi}{\bar{a}}}\right),\label{sep}\\
r_{\text{mag}} &=& \frac{\mu_1}{\sum{_{n=2}^\infty}\mu_n } \approx 
\frac{5 \pi}{\bar{a}~\text{log}(10)}\label{mag1}.
\end{eqnarray} 
In the above expression, $\theta_\infty$ is the asymptotic angular distance of the image distance, $s$ is the angular separation between $\theta_{1}$ and $\theta_{\infty}$, and $r_{\text{mag}}$ is the ratio of the flux of the first image to that of all other images.
Note that the observable $r_{\text{mag}}$ does not depend on the distance between the observer and the lens $D_{OL}$,  making it a direct probe of the spacetime geometry in the strong-field regime. By measuring these three quantities—$\theta_\infty$, $s$, and $r_{\text{mag}}$ one can, in principle, reconstruct the black hole metric parameters and distinguish between different gravitational theories.

\section{Analysis of Lensing Observables for supermassive black holes}\label{sec:anaylsis}
In this section, we apply the formalism from the previous section to numerically estimate lensing observables in SDL by treating Sgr A* and M87* as anisotropic fluid black holes in KR gravity, using the parameters inferred from EHT observations.
Using the latest astronomical observation data, the estimated mass and distance from the Earth of the M87* is given as $\left(6.5\pm0.7\right)\times{10}^9M_\odot$, and $d=16.8$ MPc \cite{Akiyama:2019}, respectively. Similarly. the estimated mass and distance of SgrA* is given as  $4_{-0.6}^{+1.1}\times{10}^6M_\odot$, and $d=8.15\pm0.15$ KPc \cite{Akiyama:2022}.

We compare the relativistic image positions $\theta_\infty$ and lensing observables, $s$ and $r_{mag}$, for an anisotropic fluid black hole in KR gravity with those for the Schwarzschild black hole. The results are summarized in Table \ref{Tableobservables}. We see that as we increase the KR coupling parameter $\ell$, the observable $\theta_\infty$ and $s$ decrease while $r_{\text{mag}}$ increases. Furthermore, the deviations in these quantities with respect to changes in $\ell$ and $K$ are more pronounced for the dust case ($w_2=0$) than for the radiation case ($w_2=1/3$). However, it is important to note that these strong deflection observables cannot be defined for the $w_2 = -1/2$ case (dark energy-like).
\begin{table*}[tbh!]
\caption{Numerical estimation of strong lensing observables for supermassive black holes Sgr A* and M87*, as an anisotropic fluid black hole in KR gravity. We compare these observables with those for Schwarzschild black holes.}\label{Tableobservables}
\begin{ruledtabular}
\begin{tabular}{c c c c c c c c}

\multicolumn{1}{c}{} & \multicolumn{1}{c}{} & \multicolumn{1}{c}{} & \multicolumn{2}{c}{Sgr A*} & \multicolumn{2}{c}{M87*} & \multicolumn{1}{c}{} \\

$w_2$ & $\ell$&$K$ & $\theta_{\rm \infty}(\mu{\rm as})$ & $s(\mu{\rm as})$ & $\theta_{\rm \infty}(\mu{\rm as})$ & $s(\mu{\rm as})$ &$r_{\rm mag}$ \\
\hline
\textbf{(GR)} & 0 & 0.0 & 26.39 & 0.32 & 19.33 & 0.233 & 6.24 \\
\hline
0 (Dust) & 0.2 & 0.0  & 18.840 & 0.00775  & 14.155 & 0.00582& 7.627 \\
&0.2 & 0.2 &  26.330 & 0.03295 & 19.782 & 0.02476 & 6.822 \\
&0.2 & 0.4 &  40.537 & 0.17933  & 30.456 & 0.13474& 5.908 \\
&0.2 & 0.6 & 74.472 & 1.4732 & 55.952 & 1.1069 & 4.824 \\
&0.2 & 0.8 &  210.64 & 29.345  & 158.26 & 22.047& 3.411 \\

&0.4 & 0.0 &  12.237 & 0.00099  & 9.194 & 0.00074& 8.807 \\
&0.4 & 0.2 & 17.102 & 0.00498  & 12.849 & 0.00374& 7.877 \\
&0.4 & 0.4 &  26.330 & 0.03295  & 19.782 & 0.02476& 6.822 \\
&0.4 & 0.6 &  48.371 & 0.34129  & 36.342 & 0.25641 & 5.570\\
&0.4 & 0.8 & 136.81 & 3.939 & 102.79 & 6.9078 & 9.1943 \\

&0.6 & 0.0 & 6.661 & $3.49\times10^{-5}$  & 5.005 & $2.62\times10^{-5}$ & 10.786\\
&0.6 & 0.2 &  9.309 & 0.000235 & 6.994 & 0.000176 & 9.648 \\
&0.6 & 0.4 & 14.332 & 0.002157  & 10.768 & 0.001621& 8.355 \\
&0.6 & 0.6 & 26.330 & 0.03295  & 19.782 & 0.02476& 6.822 \\
&0.6 & 0.8 & 74.472 & 1.4732 & 4.824 & 55.952 & 1.1069 \\

\hline
1/3 (Radiation) & 0.2 &0.0& 18.8401 & 0.00775148 & 14.1549 & 0.00582381& 7.6271 \\
& 0.2& 0.2 & 21.6894 & 0.0101109 & 16.2956 &0.0075965 & 7.20685 \\
& 0.2& 0.4 & 24.8362 & 0.0149536 & 18.6598 & 0.0112349 & 6.756 \\
& 0.2& 0.6 & 28.3439 & 0.0233588 & 21.2952 & 0.0175498 & 6.2867 \\
& 0.2& 0.8 & 32.2611 & 0.0376766 &24.2383 & 0.028307 &5.80704 \\
&0.4 & 0   & 12.237 & 0.000986  & 9.19  & 0.00074& 8.807  \\
&0.4 & 0.2 & 14.314 & 0.00165   & 10.754 & 0.00124  & 7.9443\\
&0.4 & 0.4 & 16.97 & 0.0049  & 12.75 & 0.0037 & 6.738\\
&0.4 & 0.6 & 20.509  & 0.0221 & 15.409 & 0.0166  & 5.218\\
&0.4 & 0.8 & 25.186 & 0.1341  & 18.923  & 0.1007   & 3.209\\
&0.6 & 0   & 6.6609  & $3.48\times10^{-5}$  & 5.0045  & $2.61\times10^{-5}$& 10.78 \\
&0.6 & 0.2 & 8.238  & $9.74\times10^{-5}$   & 6.18 & $7.320\times10^{-5}$& 9.354 \\
&0.6 & 0.4 & 11.013& 0.00602   & 8.27  & 0.00452 & 5.331 \\
\end{tabular}
\end{ruledtabular}
\end{table*}

\subsection{Constraints from EHT}
The EHT campaign revealed a bright, asymmetric emission ring around M87 with an angular diameter $\theta_{sh}=42\pm 3~\mu$as, exhibiting a central brightness depression—the characteristic shadow signature—and constraining the ring's fractional width to $<0.5$ \citep{EventHorizonTelescope:2019dse,EventHorizonTelescope:2019pgp,EventHorizonTelescope:2019ggy}. Subsequent analysis of Sgr A* from the same observing campaign, released in 2022, similarly confirmed a ring-like structure with a diameter of $51.8\pm2.3~\mu$as \cite{EventHorizonTelescope:2022wkp}. By combining multiple imaging techniques—including EHT Imaging, SMILI, and DIFMAP—the envelope of $1\sigma$ for the angular diameter of Sgr A* shadow is constrained to $\theta_{sh} = 48.7 \pm 7~\mu$as \cite{EventHorizonTelescope:2022xqj}. Despite M87* being approximately 1500 times more massive and 2000 times more distant than Sgr A*, their shadow diameters appear remarkably similar in the sky, making them ideal laboratories for testing gravity theories \cite{Afrin:2021wlj,Zakharov:2022gwk,Vachher:2024ait,KumarWalia:2022aop,
Ghosh:2022kit,Kumar:2023jgh,Islam:2022wck,Vachher:2024ezs}

\begin{figure}[!th]
    \centering
    \includegraphics[scale=.75]{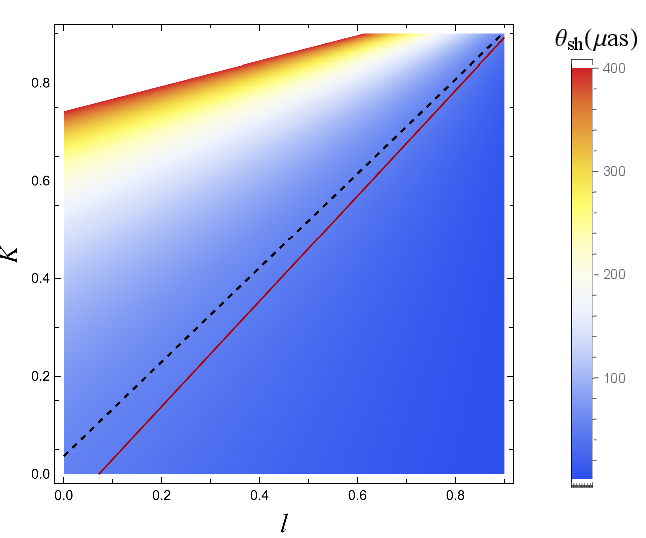}
    \includegraphics[scale=.75]{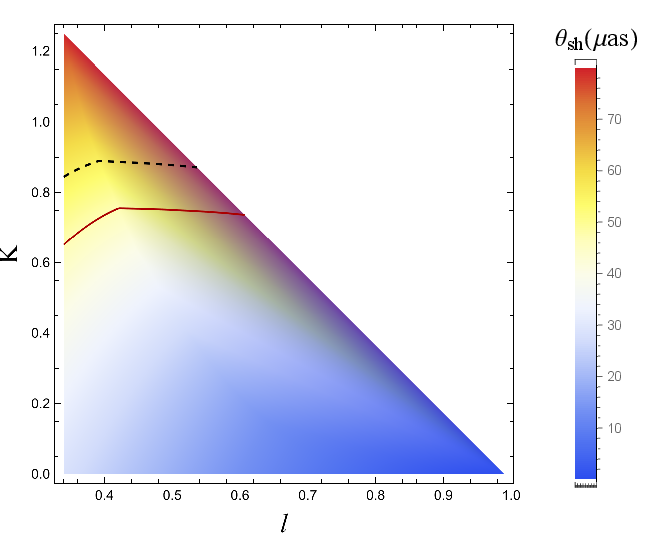}
\caption{Shadow angular diameter $\theta_{\text{sh}}(=2\theta_{\rm \infty})$ as a function of parameters $\ell$ and $K$ for Dust $w_2=0~(\text{Left})$ and Radiation $w_2=1/3~(\text{Right})$, when Sgr A* is modelled as anisotropic fluid black hole in KR gravity. The dashed black and solid lines correspond to $\theta_{\text{sh}}=55.6$ and $\theta_{\text{sh}}=41.7$, respectively. The region between these lines satisfies the Sgr A* shadow $1\sigma$ bound.}
\label{SgrAparameter}
\end{figure}

Taking the angular radius of the image position ($\theta_\infty$) as the angular size of the black hole shadow, the shadow diameter is defined as $\theta_{\text{sh}} = 2\theta_{\rm \infty}$. By modelling M87* and Sgr A* as an anisotropic fluid black hole within the Kalb-Ramond gravity framework, we can then place observational constraints on the deviation parameters $\ell$ and $K$ for different values of the equation-of-state parameter $w_2$. This is achieved by requiring that the theoretically predicted shadow diameter falls within the $1\sigma$ observational bounds reported by the EHT for each supermassive black hole.
\paragraph{Constraints from Sgr A*:}  The observed average bounds for the shadow size of Sgr A * $\theta_{\text{sh}} \in (46.9, 50)~\mu$as and the full $1\sigma$ interval as $\in$ $(41.7,55.6)~\mu$as \cite{EventHorizonTelescope:2022xqj} in Fig.~\ref{SgrAparameter}. The dashed black and solid red lines correspond to $\theta_{\text{sh}}=55.6\mu$as and $\theta_{\text{sh}}=41.7\mu$as, respectively. For dust ($w_2=0$), the $1\sigma$ bound is given as- $0\le \ell\le0.065$ and $0\le K\le0.04$, while for radiation ($w_2=1/3$), bound is- $0.65\le K\le0.85$ and no constraint on the parameter $\ell$. Within this parameter range, the anisotropic fluid black hole in KR gravity is consistent with observations of the Sgr A* black hole shadow from the EHT.  

\begin{figure}[!th]
    \centering
    \includegraphics[scale=.75]{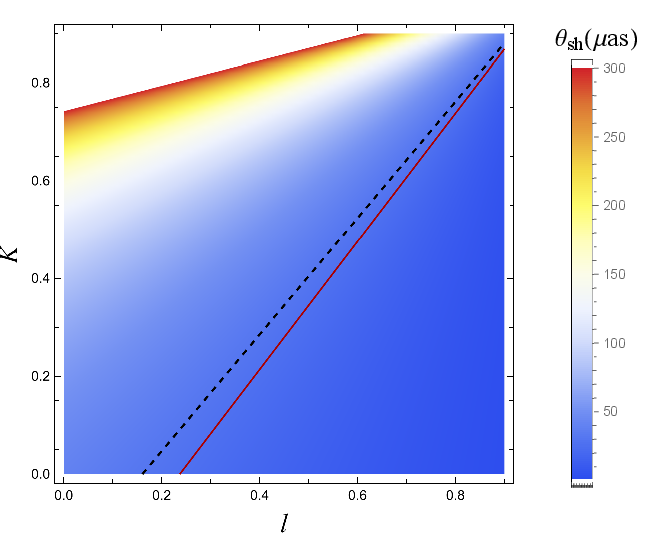}
    \includegraphics[scale=.75]{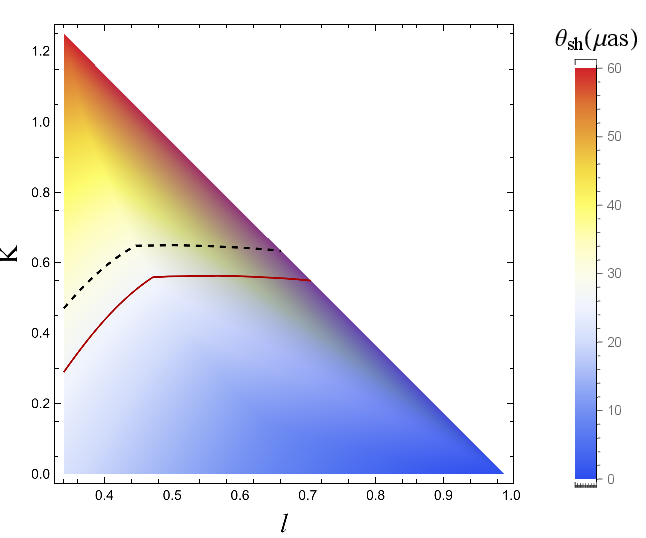}
    \caption{Shadow angular diameter $\theta_{\text{sh}}(=2\theta_{\rm \infty})$ as a function of parameters $\ell$ and $K$ for Dust $w_2=0~(\text{Left})$ and Radiation $w_2=1/3~(\text{Right})$, when M87* is modelled as anisotropic fluid black hole in KR gravity. Here, the dashed black and solid lines correspond to $\theta_{\text{sh}}=40.5$ and $\theta_{\text{sh}}=35.1$. The region within this line satisfies the M87* shadow $1\sigma$ bound.}\label{M87parameter}
\end{figure}

\paragraph{Constraints from  M87*:} In Fig.~\ref{M87parameter}, the angular diameter $\theta_{\rm sh}$ is shown as a function of $\ell$ and $K$ for the M87* black hole. Given the offset of $\approx 10\%$ between the emission ring and the angular shadow diameter, we get the angular diameter of shadow to be within the range $\in (35.1,40.5)~\mu$as \cite{Banerjee:2022bxg,EventHorizonTelescope:2019dse,Ali:2024ssf} with an error of $\pm2.7\%$ incorporating both measurement uncertainty and potential offset. Figure~\ref{M87parameter}, the angular diameter $\theta_{\rm sh}$ for the for an anisotropic fluid black hole in KR gravity as M87* where dashed black and solid lines correspond to $\theta_{\text{sh}}=40.5$ and $\theta_{\text{sh}}=35.1$, respectively. In Fig.~\ref{M87parameter}, the angular diameter $\theta_{\rm sh}$ is shown as a function of $\kappa\eta^2$ and $\gamma$ for the M87* black hole. The bounds for dust ($w_2=0$) are $0.15<\ell<0.23$ and unbound for $K$, while for radiation ($w_2=1/3$), the bounds are $0.29<K<0.45$ and unbound for $\ell$.
\section{Conclusion}\label{sec:conlusion}
In this paper, we have successfully constructed and analyzed a novel class of exact black hole solutions within a framework where gravity is non-minimally coupled to a background KR field and immersed in an anisotropic fluid. The presence of a non-zero vacuum expectation value of the KR field triggers spontaneous Lorentz symmetry breaking, which, when combined with the anisotropic matter distribution, profoundly modifies the underlying spacetime geometry. Our comprehensive analysis yields several significant physical insights. Firstly, we found that the spacetime admits a rich horizon structure dictated by the KR coupling parameter $\ell$ and the fluid density parameter $K$. We established that the asymptotic behaviour and the satisfaction of the null, weak, strong, and dominant energy conditions depend strictly on the equation-of-state parameter $w_2$, where local repulsive behaviours (dark energy-like) induce severe violations of the strong energy condition while simultaneously extending the spatial influence of the black hole. 

Furthermore, the critical parameters governing null geodesics, namely the photon sphere and the shadow radius, were derived analytically and explored numerically. We demonstrated that dark energy-like configurations ($w_2 = -1/2$) drastically alter the effective potential barrier, whereas dust ($w_2=0$) and radiation ($w_2=1/3$) backgrounds produce localised corrections to the Reissner--Nordstr\"om-like profile. We also analyzed the weak gravitational lensing in this spacetime using the Gauss–Bonnet method applied to the optical metric, which remains valid despite the non-asymptotically flat nature induced by the KR field and the surrounding anisotropic matter. The resulting deflection angle receives, in addition to the Schwarzschild-like term, corrections controlled by the fluid parameter $K$, the KR coupling $\ell$, and the equation-of-state parameter $w_2$. In all cases considered, the deflection angle decreases with the impact parameter, while larger values of $K$ and $\ell$ enhance the bending, with the strongest sensitivity observed in the DE-like background.

Finally, we extended our optical analysis to the strong-field regime to extract measurable lensing observables, including the asymptotic angular position, image separation, and magnification ratio. Applying our model to the supermassive black holes Sgr A$^*$ and M87$^*$, we revealed quantifiable deviations from general relativity. 
For Sgr A*, the asymptotic angular size of the image $\theta_\infty$ drops from $26.29\,\mu\text{as}$ for Schwarzschild to $18.84\,\mu\text{as}$ for $\ell=0.2$ and $K=0$ i.e. in case of Schwarzschild BH solution in KR gravity, and it goes upto $210.6\,\mu\text{as}$ for $K=0.8$ for dust ($w_2=0$) but for it only increase upto $32.26\,\mu\text{as}$ for radiation ($w_2=1/3$). The angular separation $s$ shows a similar trend, with separation increasing with parameter $K$ but decreasing with $\ell$. The effect of parameters are must more prominent for the dust background compared to the radiation. By comparing the theoretically predicted shadow angular diameter \(\theta_{\text{sh}} = 2\theta_\infty\) with EHT observational bounds for Sgr\,A* and M87*, we placed novel constraints on the KR field and anisotropic fluid parameters. For M87*, the 1-\(\sigma\) allowed regions are $0.15< \ell < 0.23$ and no constraint on $K$ for dust, while $0.29< K< 0.45$ and no constraint on $\ell$ for radiation. For Sgr\,A*, we get a tighter constraint of \(0 < K < 0.04\) and \(0 < \ell < 0.065\) for dust while $0.65\le K \le0.85$ and no constraint on $\ell$ for radiation.

Ultimately, this study bridges fundamental high-energy physics with observable astrophysics. The distinct lensing signatures and shadow modifications characterized here provide a concrete theoretical baseline for testing modified gravity and detecting anisotropic matter distributions, such as dark matter halos or scalar condensates, using high-resolution interferometric data.

The family of solutions derived here provides a robust theoretical foundation for exploring the strong-field phenomenology of KR gravity and its interplay with environmental anisotropy. A primary future direction is to extend these static metrics to stationary and rotating configurations, a necessary step for direct comparison with current and next-generation Event Horizon Telescope (ngEHT) data. For such rotating counterparts, the KR coupling $\ell$ and fluid parameter $K$ are expected to induce measurable asymmetries in the shadow morphology and the displacement of the photon ring, potentially providing a unique signature of Lorentz symmetry breaking. Furthermore, the dynamical stability and gravitational-wave signatures of these spacetimes can be probed through their Quasinormal Mode (QNM) spectra. The modifications to the effective potential introduced by the KR field are likely to yield distinctive shifts in the damping rates and oscillation frequencies of the $\ell=2$ fundamental mode, providing a pathway for verification by future space-based detectors such as  LISA. Additionally, these solutions provide a framework for modelling quasi-periodic oscillations (QPOs) in the X-ray spectra of accreting black hole binaries.

\begin{appendix}
\section[Derivation of Approximate Root for Small K]{Derivation of the Approximate Root Eq. (\ref{phton12}) for Small \texorpdfstring{$K$}{K}}

We consider the equation
\begin{equation}\label{A1}
\bigl[1-\ell + K(\ell-\tfrac12) r^{1/(1-\ell)}\bigr] r - 3(1-\ell)^2 M = 0, 
\end{equation}
with $-1 < \ell < 1$ and $K$ treated as a small parameter.
Let
\[
A = 1-\ell > 0,\qquad B = \ell-\tfrac12,\qquad 
C = \frac{K B}{A},\qquad D = 3A M.
\]
The equation becomes
\begin{equation}\label{A2}
r + C\, r^{1 + \frac{1}{A}} - D = 0,  
\end{equation}
where we note
\[
1 + \frac{1}{A} =  \frac{2-\ell}{1-\ell} \equiv n.
\]
For $-1 < \ell < 1$, we have $n > 0$.
We expand the root $r(K)$ as
\begin{equation}   
\label{rk}
r(K) = r_0 + r_1 K + r_2 K^2 + \mathcal{O}(K^3).
\end{equation}

Substituting into \eqref{A2} gives
\begin{equation}\label{A3}
r_0 + r_1 K + r_2 K^2 + \cdots \;+\; \frac{B}{A} K \bigl(r_0 + r_1 K + \cdots\bigr)^n - D = 0.   
\end{equation}
Setting $K=0$ in \eqref{A3} yields
\begin{equation}\label{A4}
r_0 - D = 0 \quad\Longrightarrow\quad {r_0 = D = 3(1-\ell)M}.
\end{equation}
Expand $(r_0 + r_1 K + \cdots)^n = r_0^n + n r_0^{n-1} r_1 K + \mathcal{O}(K^2)$.
The $\mathcal{O}(K)$ part of \eqref{rk} is
\[
r_1 K + \frac{B}{A} K r_0^n = 0,
\]
so

\begin{equation}    
\label{A5}
r_1 = -\frac{B}{A}\, r_0^n.
\end{equation}
Since $B = \ell-\tfrac12$, $A = 1-\ell$, and $r_0 = 3(1-\ell)M$,
\[
r_0^n = \bigl[3(1-\ell)M\bigr]^{\frac{2-\ell}{1-\ell}}.
\]
Hence
\begin{equation}   
\label{A6}
r_1 = -\frac{\ell-\frac{1}{2}}{1-\ell}\,
\bigl[3(1-\ell)M\bigr]^{\frac{2-\ell}{1-\ell}}
\end{equation}
\begin{equation}   
\label{A7}
r(K) \approx 3(1-\ell)M 
\;-\; \frac{\ell-\frac{1}{2}}{1-\ell}\,
\bigl[3(1-\ell)M\bigr]^{\frac{2-\ell}{1-\ell}} K
\;+\; \mathcal{O}(K^2)
\end{equation}

For the Special case $\ell = 1/2$, we have $B = 0$, so $r_1 = 0$ and the term proportional to $K$ vanishes identically.  
Equation \eqref{A2} reduces to $r - D = 0$, giving the exact solution
\[
r = 3(1-\tfrac12)M = \frac{3M}{2},
\]
independent of $K$, in agreement with \eqref{A7}.
\end{appendix}


\bibliographystyle{ieeetr}  
\bibliography{1references.bib}

\end{document}